\documentclass[%
superscriptaddress,
 amsmath,amssymb,
 aps,
prb,
 reprint,
floatfix,
]{revtex4-2}
\usepackage{graphicx,xcolor}
\definecolor{darkblue}{RGB}{0,0,150}
\definecolor{nightblue}{RGB}{0,0,100}

\usepackage{mathrsfs,dsfont,mathtools}

\usepackage{ulem}

\usepackage{dcolumn}
\usepackage{bm}
\usepackage[
colorlinks,
citecolor=darkblue,
linkcolor=darkblue,
urlcolor=nightblue]{hyperref}

\usepackage[english]{babel}
\usepackage[babel,kerning=true,spacing=true]{microtype}

\usepackage{feynmp-auto}

\AtBeginDocument{\renewcommand{\natexlab}[1]{#1}}
\begin{document}

\author{Daniel Kaplan}\thanks{These authors contributed equally.}
\affiliation{Department of Physics and Astronomy, Rutgers University, Piscataway, NJ 08854, USA}
\affiliation{Center for Materials Theory, Rutgers University, Piscataway, NJ 08854, USA}
\author{Alexander C. Tyner}\thanks{These authors contributed equally.}
\affiliation{Nordita, KTH Royal Institute of Technology and Stockholm University 106 91 Stockholm, Sweden}
\affiliation{Department of Physics, University of Connecticut, Storrs, Connecticut 06269, USA}
\author{Eva Y. Andrei}
\affiliation{Department of Physics and Astronomy, Rutgers University, Piscataway, NJ 08854, USA}
\author{J. H. Pixley}
\affiliation{Department of Physics and Astronomy, Rutgers University, Piscataway, NJ 08854, USA}
\affiliation{Center for Materials Theory, Rutgers University, Piscataway, NJ 08854, USA}
\affiliation{Center for Computational Quantum Physics, Flatiron Institute, 162 5th Avenue, New York, NY 10010}
\title{Machine learning assisted high throughput prediction of moir\'e materials}
\date{\today}
\begin{abstract}
The world of 2D materials is rapidly expanding with new discoveries of stackable and twistable layered systems composed of lattices of different symmetries, orbital character, and structural motifs. Often, however, it is not clear {\sl a priori} whether a pair of monolayers twisted at a small angle will exhibit correlated or interaction-driven phenomena. The computational cost to make accurate predictions of the single particle states is significant, as small twists require very large unit cells, easily encompassing 10,000 atoms, and therefore implementing a high throughput prediction has been out of reach. 
Here we show a path to overcome this challenge by introducing a machine learning (ML) based methodology that {\sl efficiently} estimates the twisted interlayer tunneling at arbitrarily low twist angles through the local-configuration based approach that enables interpolating the local stacking for a range of twist angles using a random forest regression algorithm.

We leverage the kernel polynomial method to compute the density of states (DOS) on large real space graphs by reconstructing a lattice model of the twisted bilayer with the ML fitted hoppings. For twisted bilayer graphene (TBG), we show the ability of the method to resolve the magic angle DOS at a substantial improvement in computational time. We use this new technique to scan through the database of stable 2D monolayers (MC2D) and reveal new twistable candidates across the five possible points groups in two-dimensions with a large DOS near the Fermi energy, with potentially exciting interacting physics to be probed in future experiments. 
\end{abstract}

\maketitle

\section{Introduction}
The design and control of two-dimensional materials has reached an unprecedented level~\cite{Andrei2021,Mak2022,Pixley2025,Bernevig2025}. With the discovery of correlated insulators~\cite{Cao2018b} and superconductors~\cite{Cao2018} in twisted bilayer graphene~\cite{Li2010}, the field quickly moved on to twisting and stacking a hand-full of transition metal dichalcogenides~\cite{Mak2022,Ghiotto2021,Zhao2023}, and a cuprate van der Waals superconductor~\cite{ZhaoS2023}. Similar physics has been identified in rhombohedral graphene~\cite{CastroNeto2009,Lu2024,Han2025} devices without the requirement of twisting individual layers. The landscape of potential exfoliate-able materials is vast, with a total number of potential van der Waals materials totaling in the thousands~\cite{Mounet2018,campi2023expansion}. The resulting number of possible stacking combinations between different materials is factorial in that number. It is therefore essential that we develop new theoretical tools to quickly diagnose which materials are the most interesting to twist and stack, and which are not useful in this regard.

\begin{figure*}
    \centering
    \includegraphics[width=17cm]{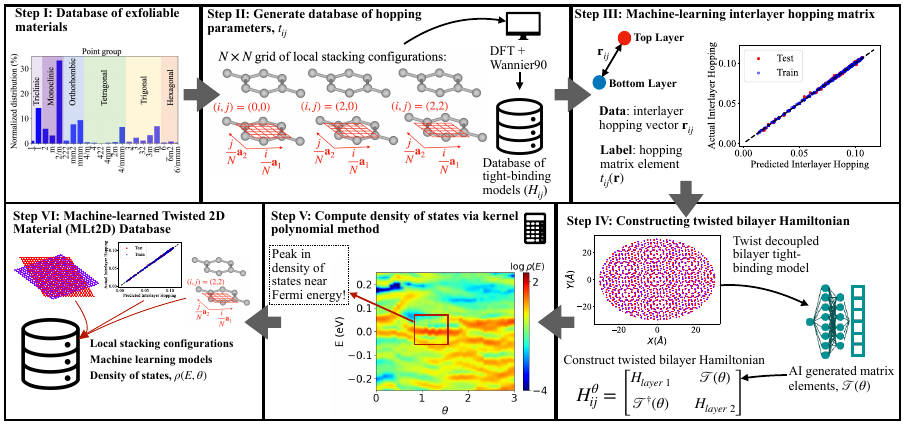}
    \caption{\textbf{Generation of the Machine-learned Twisted 2D Material (MLt2D) Database:} (Step I) Workflow for construction of the MLt2D database begins with exfoliable two-dimensional materials from the Materials Cloud two-dimensional structure database~\cite{mounet2018two,campi2023expansion}.  (Step II) Bilayers of the selected materials are formed and density functional theory computations are performed for a $10 \times 10$ of local stacking configurations, resulting in a dataset of interlayer hopping matrix elements. (Step III) The interlayer hopping matrix elements are used to train and ensemble of random forest machine learning (ML) networks. (Step IV) A tight-binding model of the twisted bilayer is formed using the ML networks to populate the interlayer matrix elements at arbitrary twist. (Step V) The density of states is computed via the kernel polynomial method. Here the DOS is obtained on TBG showing the appearance of the flat band near the magic-angle of $\theta\approx 1.1^\circ$ (Step VI) Local stacking configurations, machine learning model and density of states data will be cataloged in a future MLt2D database.}
    \label{fig:Fig1}
\end{figure*}

The current high-throughput approach to this particular ``problem of choice'' is to utilize symmetry indicators and underlying  topologies of the systems band structure before it is down-folded through a moire potential~\cite{jiang20242d,Crepel2025,Lhachemi2025,Nakatsuji2025}.  
The fundamental challenge in considering the moir\'e system with a fully ab initio calculation is that the unit cell is prohibitively large~\cite{Zhang2022}, requiring on the order of 10,000 atoms (e.g., near the magic-angle of twisted bilayer graphene). As density functional theory computations scale cubically, $O(N^3)$, with the number of atoms, even a single computation requires $\sim 10^{6}$ core hours. 
\par 
As a result, approximate methods have been developed that utilize efficient and accurate ab initio calculations of the bilayer unit cell, which contains a small number of atoms, across multiple stacking configurations. Analysis of these distinct configurations can then be used to reconstruct approximately the effect of a twisted bilayer when the angle is small~\cite{Fang2016,massatt2017,Carr2017,Cances2017}. Importantly, this approach converges to the thermodynamic limit exponentially fast, as opposed to algebraically, which a conventional brute force construction would afford. However, to date, this ``local-configuration'' has required manual intervention and has yet to be systematically automated for use in a high-throughput approach, until now. 

In this work, we develop an efficient and scalable high-throughput method, laid out in Fig.~\ref{fig:Fig1}, that utilizes the ``local-configuration'' approximation~\cite{Fang2016,massatt2017,Carr2017,Cances2017} to train an ensemble of random-forest machine learning networks for efficient and accurate estimate of twisted interlayer Hamiltonian matrix elements at arbitrary twist angles. From these effective tight binding models of twisted bilayers, we compute the density of states using the Chebyshev expansion based approach, known as the kernel polynomial method (KPM). This reduces a typical simulation time from $10^{6}$ to $10^{3}$ core hours, making intractable calculations approachable for the first time.
We apply this to leading, stable 2D van der Waals materials in the Materials Cloud two-dimensional structure (MC2D) database \cite{mounet2018two,campi2023expansion} and list them in their difficulty to exfoliate. We then describe material candidates that have emerged from this search.

\section{Approach}\label{sec:2}
To construct the electronic structure of twisted bilayers we construct an effective bilayer Hamiltonian, where the dominant effect is tunneling between the two layers, which is given by
\begin{equation}\label{eq:BilayerHam}
    H= \begin{pmatrix}
        H_{\mathrm{layer \, 1}} & \mathcal{T}(\theta)
        \\
        \mathcal{T}(\theta)^{\dag} &  H_{\mathrm{layer \, 2}}
    \end{pmatrix}.
\end{equation}
Each 2D Hamiltonian $H_{\mathrm{layer \, 1/2}}$ can be well described using ab initio approaches (for the weakly correlated van der Waals materials). However, efficiently calculating the twisted interlayer tunneling $\mathcal{T}(\theta)$  represents a timely and fundamental challenge for computational materials science given the immense ($\sim10^{4}$) number of atoms in the primitive unit cell. The extraction of the electronic structure via density functional theory (DFT) for a single twisted system of this size is possible\cite{Song2019,devakul2021magic,zhong2023transferable,gong2023general,qi2025bridging}, but variation of the twist-angle and the subsequent screening of many compounds quickly becomes unrealistic. While continuum models can be constructed for each system under consideration, each is tailored to the material and do not generalize, which precludes them from being used directly for a broad search. On the other hand, prior work using machine-learning to try and overcome this challenge generated the moir\'e Hamiltonian using deep learning methods\cite{zhong2023transferable,gong2023general,qi2025bridging}.  
These works have focused primarily on twisted bilayer graphene and twisted bilayer TMD MoS$_{2}$. However, this approach requires an immense amount of training data, which has to be generated from DFT computations that are computationally expensive and therefore this approach will not be useful to scan a broad class of materials.

\subsection{Efficiently estimating the twisted interlayer tunneling}
\par
The first step for a given homo-bilayer is DFT computations using the ``local-configuration'' method introduced in Ref.~\cite{Fang2016}. Details of the density-functional calculations are provided in Methods.

In the ``local-configuration'' method, sketched in Fig.~\ref{fig:Fig1} step II, a series of small-scale DFT computations are performed using the primitive unit cell of the untwisted homo-bilayer. In each computation, the bottom-layer is shifted in-plane relative to the top-layer by $\frac{i}{20}\mathbf{a}_{1}+\frac{j}{20}\mathbf{a}_{2}$ where $\mathbf{a}_{i=1,2}$ are the in-plane lattice vectors of the primitive unit cell and $i,j\in [1,10]$. For each DFT computation performed on this $10 \times 10$ grid, the atomic positions in the out-of-plane direction are relaxed. Finally, a Wannier tight-binding (WTB) model is constructed, which precisely replicates the electronic structure for the system near the Fermi energy at each grid point using the Wannier90 software\cite{Pizzi2020}. Twisting the bottom layer by $\theta_1$ and the top layer by $\theta_2$, our goal is to estimate $t_{ij}(\bm r,\theta_1,\theta_2)$ the interlayer tunneling between locations $i$ and $j$ that are separated by the vector $\bm r$. To provide a direct parameterization of the leading terms, one can expand the tunneling in terms of the cylindrical harmonics~\cite{Fang2016}, instead we fit this function from all of the data we obtain from the ``local configuration'' approach as we now describe and show in Fig.~\ref{fig:Fig1} step III.

\begin{figure}
    \centering
    \includegraphics[width=8cm]{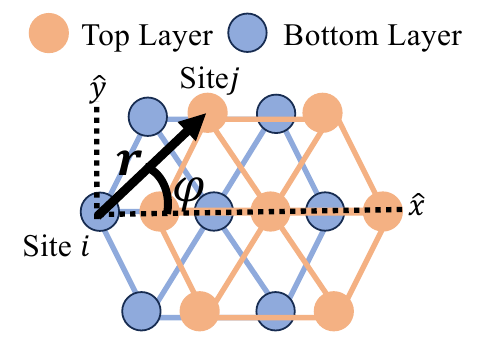}
    \caption{Schematic of relative interlayer hopping vector, $\mathbf{r}$, and corresponding in-plane angle from the x-axis, $\phi$. This vector and angle are used to label interlayer tunneling matrix elements in training of the random forest machine learning models.}
    \label{fig:input_data}
\end{figure}
\par 
Using the WTB models, we then create a dictionary containing the interlayer hopping matrix between atoms in the same unit cell. Each hopping matrix in this dictionary is labeled by the vector separating the atoms, $\mathbf{r}=\{r_{x},r_{y},r_{z} \}$, which is decomposed into two parts: the magnitude, $|\mathbf{r}|$, and the in-plane angle measured from the x-axis, $\phi=\tan^{-1}(r_{y}/r_{x})$, shown schematically in Fig.~\eqref{fig:input_data}. As we consider only a single orbital pair at a time, we can invoke the periodicity of $\phi$ to fix $\phi=0$ for the untwisted unit cell. It is then possible to define $\phi =\theta_{1}-\theta_{2}$.
Using this dictionary, we isolate all interlayer orbital pairs which, for at least a single WTB model, support interlayer hopping greater than $10^{-2}\; eV$ within the same unit cell. For each pair meeting this requirement, a random forest regression model is trained, which takes the input vector $(r,\phi)$ and predicts a value for the interlayer hopping. This process represents a pathway to automating the fitting protocol presented in Ref.~\cite{Fang2016,Carr2019,carr2020electronic}. 
We  utilize a random forest model as opposed to a deep-learning method due to the limited data set size; for each orbital pair the dataset size is given by grid over which the ``local-configuration'' method is deployed, 100  data points in our case. For this limited dataset size the random forest model provides accuracy and computational speed without overfitting, which would be challenging to avoid using a deep-learning approach.
\par 
In order to construct the Hamiltonian for the twisted bilayer, we have developed a code based on the PyBinding software package~\cite{dean_moldovan_2020_4010216}, depicted in Fig.~\ref{fig:Fig1} step IV. This code imports the untwisted bilayer in the limit of zero interlayer coupling, and subsequently twists the top layer by an arbitrary angle, $\theta$. After the twist is applied, the trained machine learning models are used to populate the interlayer hopping matrix elements for all relevant sites labeled by atom, orbital and spin. As the bilayers studied in this work are selected from a database of easily exfoliable two-dimensional materials\cite{mounet2018two,campi2023expansion}, the intralayer Hamiltonian is fixed for all twist angles (further details of the twisted bilayer Hamiltonian construction are in Methods).  
Upon generation of interlayer hopping terms for all relevant orbital/spin pairs indistinct layers, a Hamiltonian representing the twisted bilayer has been formed. This Hamiltonian represents progress in computational materials science modeling of twisted materials, as it is derived entirely from first-principles models with no fitting to a continuum model. Furthermore, the entire process can be automated with no manual intervention necessary.
\subsection{Extracting the bulk density of states}
To consider arbitrary twist angles in the search process, we have to consider the real space lattice model with open-boundary conditions, removing translational symmetry. This produces matrix representations of the Hamiltonian that easily reach sizes of $\sim 10^{6}\times10^6$ to access moir\'e physics. We therefore utilize the KPM to analyze the electronic structure through the  density of states, as shown in Fig.~\ref{fig:Fig1} step V.  In the following, we focus on the bulk density of states, which is extracted by first computing the local density of states, $\rho_{i,\tau,\sigma}(E)$ (see Methods for definition), 
where $i,\tau,\sigma$ correspond to site, orbital and spin degrees of freedom, restricting the sum over $i$ to sites in the middle of the sample. Namely within a $10 \AA$ radius. 
This is to avoid spurious contributions due to edge modes as we are considering a finite-size disk geometry. 

To produce an efficient search across this large material set, we analyze the bulk density of states in the vicinity of the Fermi energy as a function of twist angle, which can reveal flat bands, isolated minibands, and van Hove singularities.
Importantly, our approach is flexible as KPM implementations of additional quantities of interest, such as spectral functions accessible to ARPES, and the conductivity\cite{kpmchern} that can be measured in transport experiments, are available and will be the focus of future work.

\section{Results}
\label{sec:3}

We selectively search through the MC2D \cite{Mounet2018,campi2023expansion} by filtering for criteria that maximize the experimental relevance of candidates. 
\begin{figure}[!ht]
    \centering
    \includegraphics[width=0.9\columnwidth]{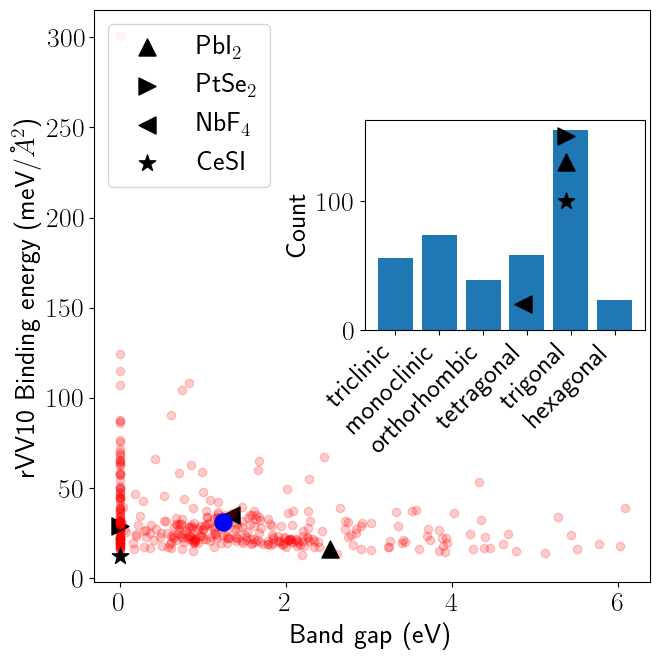}
    \caption{Distribution of crystal classes, band gaps and vdW (van der Waals)  binding energies $E_b$, denoted by red dots. We mark the four candidates we focus on with different symbols. (Inset) Total number of different materials for each point group symmetry.}
    \label{fig:distr}
\end{figure}
We exclusively filter for materials with known phonon dispersion, which \textit{do not} contain imaginary modes indicating their structural stability. We then order these by the computed rVV10 \cite{peng2016versatile} 2D binding energy $E_b$ in meV/$\AA^2$, as determined in Ref.~\cite{Mounet2018}. This guarantees that the bilayers in question can be created and importantly twisted to an arbitrary twist angle \cite{zhang2016van}. We find an assortment of materials with low $E_b$, and the stable ones with computed $E_b$ via rVV10 total 435 different materials. Of these, $E_b < 20 \textrm{meV}/\AA^2 $ are of particular interest, and of these there are 78 at present in the MC2D. In Fig.~\ref{fig:distr}(a) we plot the distribution of the stable materials across crystal classes, finding an over-representation of trigonal lattices, but a sizable contribution across all relevant crystals groups for 2D tilings. In Fig.~\ref{fig:distr}(b), we show that our selection of materials has considerable variability in both the calculated DFT band gap (horizontal axis) and vdW binding energy. The mean value is a band gap of $E_g \approx 1.1 \textrm{eV}$ and binding energy of $E_b \approx 27 \textrm{meV}/\AA^2$ (blue dot). 

\subsection{PbI$_2$}
In searching for {\textit{exfoliable}} candidates, iodides are ideal candidates \cite{mustonen2022toward} due to the inherent property of iodine volatility and instability.
\begin{figure}[!ht]
    \centering
    \includegraphics[width=1.0\columnwidth]{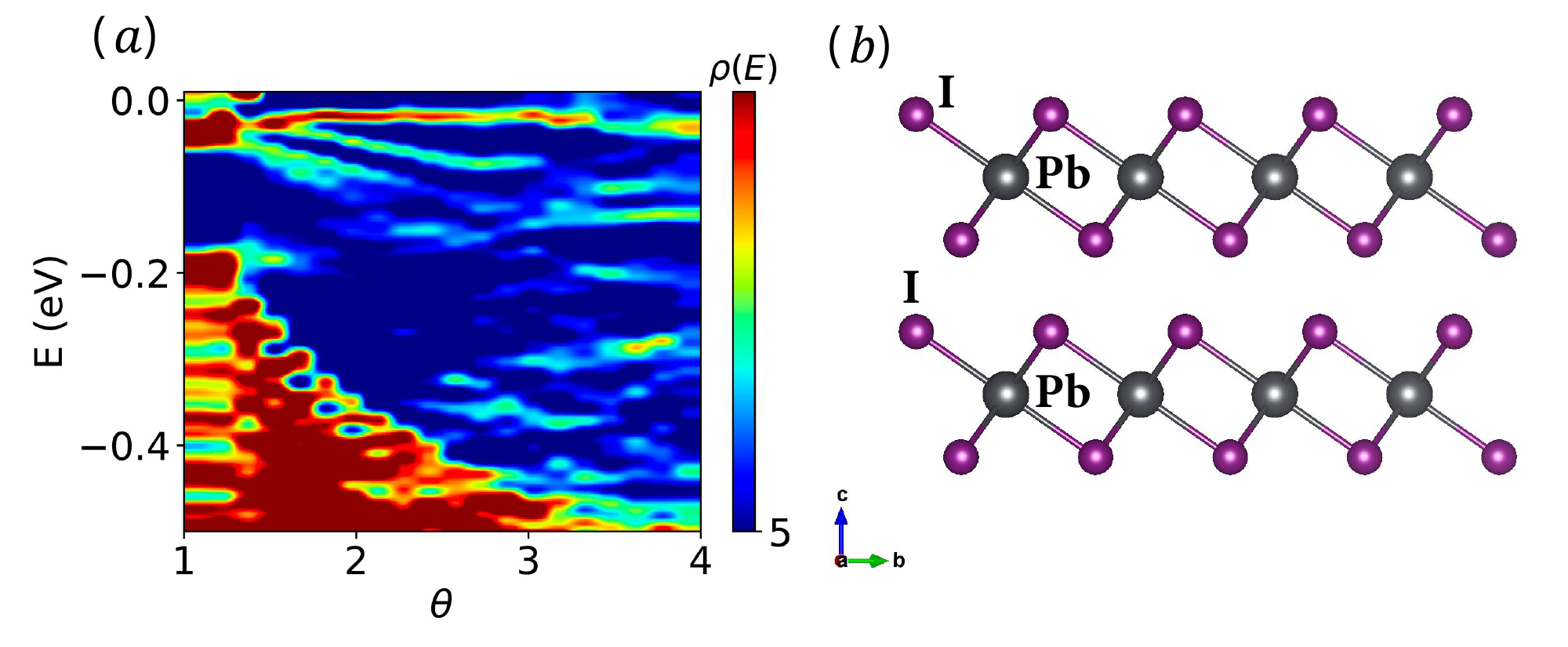}
    \caption{(a) The DOS map of twisted bilayer Pb$I_2$ as a function of energy (E) in units of eV vs the twist angle $\theta.$ We find a flat band develops around 2$^{\circ}$ near zero Fermi energy. (b) Crystal structure of the bilayer.}
    \label{fig:PbI2}
\end{figure}
This, in turn, weakens vdW like bonding between layers, especially in scenario where iodine serves as a ligand, effectively a wedge between layers.
Scanning through MC2D, we found PbI$_2$ (mc2d-183), with an a binding energy of $E_b = 16.12 \textrm{me}V/\AA^2$, only slightly above the weakest bonded material in the database, BiI$_3$ -- another iodide. This material is well known in the 3D form for its uses in perovskite solar cells \cite{beckmann2010review,xu2016formation}. In Fig.~\ref{fig:PbI2}(a), we plot the DOS as a function of twisting bilayer PbI$_2$~ Fig.~\ref{fig:PbI2}(b). The monolayer arranges in a 1T structure, making the individual layer inversion symmetric ($\bar{3}m$ point group). As Pb is a heavy element, SOC effects are expected to be especially prominent in this system. The aligned structure (Fig.~\ref{fig:PbI2}(b)) retains inversion, and therefore suppresses effects of Rashba SOC. However, upon twisting, inversion is broken  and SOC splittings become more pronounced. We find a large range of enhanced DOS for $\theta < 4 \deg$ with a collection of minibands dispersing within $E \pm 0.1 \textrm{eV}$ from the Fermi level (which is assume pinned to the top of the valence band, whose minimum is at $\Gamma$, see appendix). We suggest that this enhanced DOS would be ideal for probing strongly interacting states of matter, by doping into the band. This would combine the effects of exceptionally strong SOC (tunable by twist angle) while retaining the density of states needed for nested instabilities.
\subsection{PtSe$_2$}
Transition metal dichalcogenides have become the mainstay in the observation of strongly correlated phenomena, exotic superconductivity~\cite{Mak2022,xia2025superconductivity,guo2025superconductivity} in moir\'e materials.
PtSe$_2$ (mc2d-227) arranges in the 1T phase, similar to Pb$I_2$ (Fig.~\ref{fig:PbI2}(b)). 
\begin{figure}[!ht]
    \centering
    \includegraphics[width=0.8\columnwidth]{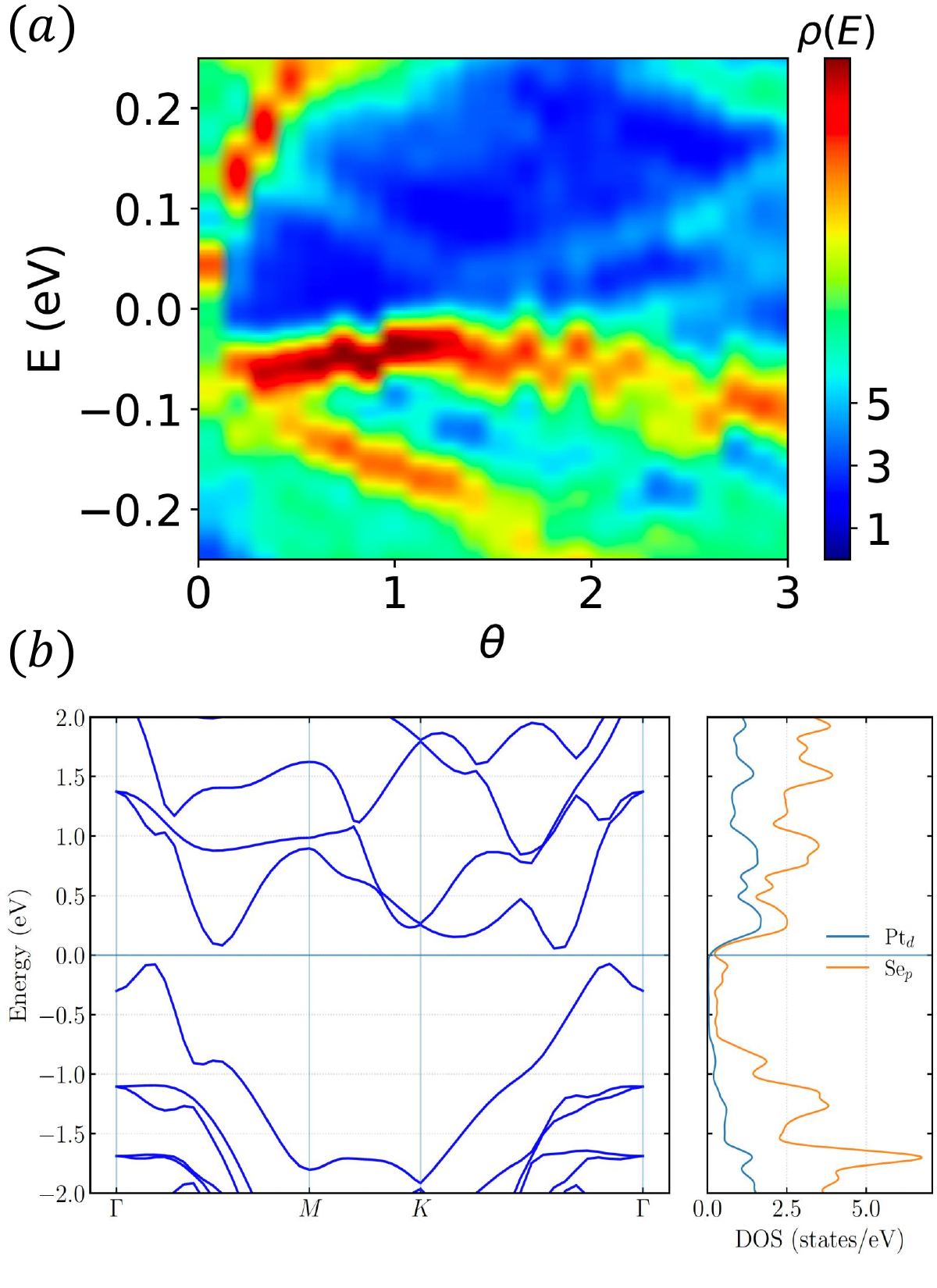}
    \caption{(a) The DOS map of twisted bilayer PtSe$_2$ as a function of energy ($E$) in units of eV vs the twist angle $\theta$ with the color marking the value of the DOS. We find a clear enhanced band emerge near a twist of $1^\circ$. (b) The band structure (left) and DOS (right) of bilayer of PtSe$_2$ at AA stacking, showing the typical band alignment. The gap is significantly modified compared to the monolayer, and the orbital character (right) shows considerable mixing between $p$, $d$ orbitals. For simplicity, spin-orbit coupling was not included for the purposes of determining the orbital composition of the DOS.}
    \label{fig:PtSe2}
\end{figure}
Using our method, we compute the DOS of the twisted bilayer, Fig.~\ref{fig:PtSe2}(a) we find a significant enhancement of the DOS around $\theta \lesssim 1 \deg$, which then disappears near $\theta  \approx 0$. We also see the emergence of minibands and fine spectral features as the twist angle evolves. These minibands then merge near $\theta \approx 2.8 \deg$ before becoming fainter.

To understand the phenomenology of the states at small twist angle, we examine the band structure of AA-stacked untwisted bilayer PtSe$_2$. We find a large density of states near $\Gamma$, but not exactly at $\Gamma$ enabling a DOS increase when doping into the bands, both for electorns and holes. We note that this system will also allow for selective orbital character, as electrons are more $d$- like, whereas holes are $p$-like. We expect this system will also exhibit rich topological features as a result of orbital mixing.
\subsection{NbF$_4$}
The search for twistable monolayers that are not trigonal is exciting enormous interest. This possibility was further confirmed in our inspection of the MC2D database (Fig.~\ref{fig:distr} (a)).
\begin{figure}[!ht]
    \centering
    \includegraphics[width=1.0\columnwidth]{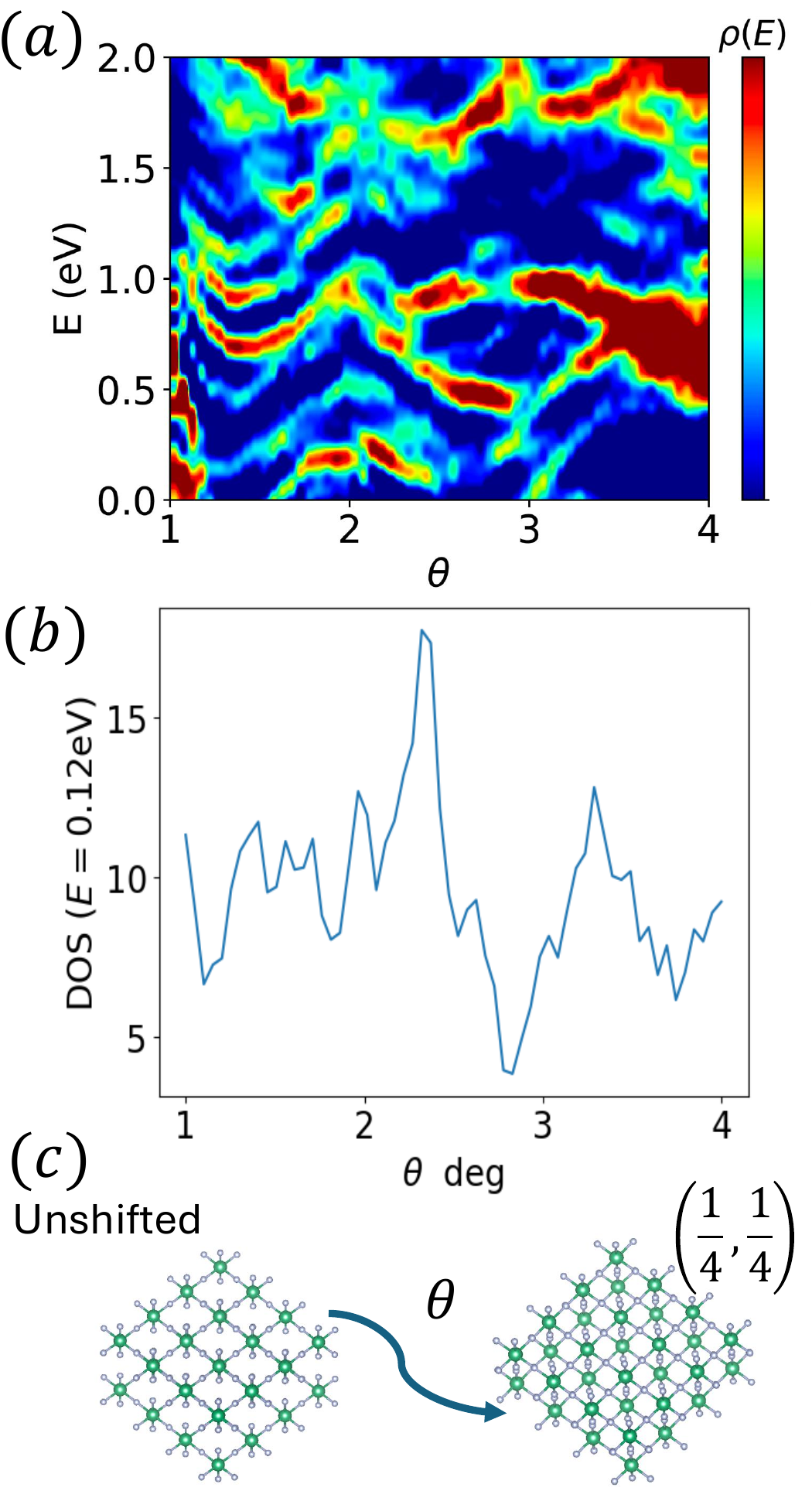}
    \caption{The DOS map of twisted bilayer NbF$_4$ (a) as a function of energy ($E$) in units of eV vs the twist angle $\theta$ with the color marking the value of the DOS. 
    (b) We show a cut at $E=0.12\textrm{eV}$ where the DOS shows distinct peaks indicating a region of potential flat band formation.
    (c) We show the nature of the square lattice at AA (zero shift) and shifted by (1/4,1/4), forming an interpenetrating lattice of Nb atoms, which will be realized by twisting the tetragonal monolayer.}
    \label{fig:NbF4}
\end{figure}
One natural choice would be the tetragonal group, and we find that the material with the lowest $E_b = 29.067 \textrm{meV}/\AA^2$ is NbF$_4$ (mc2d-139) whose monolayer arranges in the symmorphic $4/mmm$ point group. This greatly simplifies the symmetry discussion. Focusing on the conduction band, we find a large space of minibands and DOS enhancement at a variety of low twist angles $\theta \lesssim 4 \deg$, as we show in Fig.~\ref{fig:NbF4}(a). These are predominantly Nb-$d$ bands which are connected in the elemental compounds to superconductivity. The moir\'e lattice is expected to interpolate over stacking configurations which are relevant for strongly correlated physics. In Fig.~\ref{fig:NbF4}(b) we show how the AA stacked bilayer forms a square lattice. But the periodicity, and interpenetration of Nb atoms between layers can be tuned by shifting the top layer by (1/4, 1/4) (Fig.~\ref{fig:NbF4}(c)). As the twisted structure includes domains of such shifts, it is conceivable that a network of square-lattice-like islands with different motifs will be formed, potentially connecting Lieb-lattice physics and more conventional square lattice electronic properties.
\section{Discussion and conclusions}
\label{sec:4}
In this work, we introduce a new computational scheme for efficiently screening large datasets of materials, such as MC2D for targeted properties as moir\'e materials, which can be accessed with real-space computational method such as the KPM. We combed through the MC2D database by focusing on the experimental relevance of candidates, filtering for stable materials, and ranking them by exfoliability as controlled by the vdW energy of stacking. In this process, we validated our method on TBG (Fig.~\ref{fig:Fig1}(inset)), and revealed that our approach correctly captures the DOS enhancement at $\theta \approx 1.1 \deg$, which is the magic angle. Following this check, we carried out calculations for a variety of materials and presented PbI$_2$, PtSe$_2$, and NbF$_4$ as leading candidates that present a large DOS near the Fermi level upon twising. One of the advances we are able to report is the significant reduction in computational time, in carrying out these scans: from the perspective of ordinary DFT calculations, an order-of-magnitude of core-hours was saved in carrying out our approach, with satisfactory (and improvable) reliability and accuracy. We stress that a principle outstanding challenge is the effective {\textit{initial screening}} of candidates for experiment, which our work has aimed to alleviate; having selected likely interesting materials, an accurate and extensive DFT study can be carried out and for the present candidates will be the subject of future work.

One limitation of our approach is its reliance on the existence of certain data in the MC2D database; chiefly, phonon calculations which are numerically intensive. We highlight a candidate that was not included in our initial scan, due to the absence of phonon dispersion -- CeSI, (mc2d-2196) -- which nevertheless is intriguing from a variety of perspectives:  it is metallic, with (within DFT) Ce $f$ levels near $E_F$;  it has a low exfoliation energy (taking DF2-CO9 for the vdW correction) of $E_b \approx 10.42 \textrm{meV}/\AA^2$, one of the lowest in MC2D; and its sister compound, CeSiI has recently been realized in experiment~\cite{posey2024two}.
This suggests that relaxing some of the stringent constraints we placed when scanning MC2D could reveal a plethora of rich, new, experimentally viable candidates for low-dimensional physics with profoundly exciting new twistable materials.

\begin{acknowledgements}
We sincerely thank Naoto Nakatsuji, Jennifer Cano and Valentin Cr\'epel for sharing, in advance, a preprint of Ref.~\cite{Nakatsuji2025}. We are obliged to M. Luskin for fruitful discussions and related collaboration. DK is supported by the Abrahams Postdoctoral Fellowship
of the Center for Materials Theory at Rutgers University (D.~K.) and the Zuckerman STEM Fellowship (D.~K.), DOE-FG02-99ER45742 (E.~Y.~A.), the Gordon and Betty Moore Foundation EPiQS initiative  GBMF9453 (E.~Y.~A.), and NSF Grant No.~DMR-2515945 (J.~H.~P.). This work was  performed  in part at the Aspen Center for Physics, which is supported by National Science Foundation grant PHY-2210452 (D.~K., A.~T.~, J.~H.~P.) and as well as at the Kavli Institute of Theoretical Physics that is supported in part by the National Science Foundation under Grants No.~NSF PHY-1748958 and PHY-2309135 (D.~K., E.~A., J.~H.~P.). NORDITA is supported in part by NordForsk. A.T. acknowledges computational resources provided by the National Academic Infrastructure for Supercomputing in Sweden (NAISS), partially funded by the Swedish Research Council through grant agreement no. 2022-06725.  

\end{acknowledgements}
\bibliography{main.bbl}

\begin{thebibliography}{50}%
\makeatletter
\providecommand \@ifxundefined [1]{%
 \@ifx{#1\undefined}
}%
\providecommand \@ifnum [1]{%
 \ifnum #1\expandafter \@firstoftwo
 \else \expandafter \@secondoftwo
 \fi
}%
\providecommand \@ifx [1]{%
 \ifx #1\expandafter \@firstoftwo
 \else \expandafter \@secondoftwo
 \fi
}%
\providecommand \natexlab [1]{#1}%
\providecommand \enquote  [1]{``#1''}%
\providecommand \bibnamefont  [1]{#1}%
\providecommand \bibfnamefont [1]{#1}%
\providecommand \citenamefont [1]{#1}%
\providecommand \href@noop [0]{\@secondoftwo}%
\providecommand \href [0]{\begingroup \@sanitize@url \@href}%
\providecommand \@href[1]{\@@startlink{#1}\@@href}%
\providecommand \@@href[1]{\endgroup#1\@@endlink}%
\providecommand \@sanitize@url [0]{\catcode `\\12\catcode `\$12\catcode
  `\&12\catcode `\#12\catcode `\^12\catcode `\_12\catcode `\%12\relax}%
\providecommand \@@startlink[1]{}%
\providecommand \@@endlink[0]{}%
\providecommand \url  [0]{\begingroup\@sanitize@url \@url }%
\providecommand \@url [1]{\endgroup\@href {#1}{\urlprefix }}%
\providecommand \urlprefix  [0]{URL }%
\providecommand \Eprint [0]{\href }%
\providecommand \doibase [0]{https://doi.org/}%
\providecommand \selectlanguage [0]{\@gobble}%
\providecommand \bibinfo  [0]{\@secondoftwo}%
\providecommand \bibfield  [0]{\@secondoftwo}%
\providecommand \translation [1]{[#1]}%
\providecommand \BibitemOpen [0]{}%
\providecommand \bibitemStop [0]{}%
\providecommand \bibitemNoStop [0]{.\EOS\space}%
\providecommand \EOS [0]{\spacefactor3000\relax}%
\providecommand \BibitemShut  [1]{\csname bibitem#1\endcsname}%
\let\auto@bib@innerbib\@empty
\bibitem [{\citenamefont {Andrei}\ \emph {et~al.}(2021)\citenamefont {Andrei},
  \citenamefont {Efetov}, \citenamefont {Jarillo-Herrero}, \citenamefont
  {MacDonald}, \citenamefont {Mak}, \citenamefont {Senthil}, \citenamefont
  {Tutuc}, \citenamefont {Yazdani},\ and\ \citenamefont {Young}}]{Andrei2021}%
  \BibitemOpen
  \bibfield  {author} {\bibinfo {author} {\bibfnamefont {E.~Y.}\ \bibnamefont
  {Andrei}}, \bibinfo {author} {\bibfnamefont {D.~K.}\ \bibnamefont {Efetov}},
  \bibinfo {author} {\bibfnamefont {P.}~\bibnamefont {Jarillo-Herrero}},
  \bibinfo {author} {\bibfnamefont {A.~H.}\ \bibnamefont {MacDonald}}, \bibinfo
  {author} {\bibfnamefont {K.~F.}\ \bibnamefont {Mak}}, \bibinfo {author}
  {\bibfnamefont {T.}~\bibnamefont {Senthil}}, \bibinfo {author} {\bibfnamefont
  {E.}~\bibnamefont {Tutuc}}, \bibinfo {author} {\bibfnamefont
  {A.}~\bibnamefont {Yazdani}},\ and\ \bibinfo {author} {\bibfnamefont {A.~F.}\
  \bibnamefont {Young}},\ }\bibfield  {title} {\bibinfo {title} {The marvels of
  moir{\'e} materials},\ }\href@noop {} {\bibfield  {journal} {\bibinfo
  {journal} {Nature Reviews Materials}\ }\textbf {\bibinfo {volume} {6}},\
  \bibinfo {pages} {201} (\bibinfo {year} {2021})}\BibitemShut {NoStop}%
\bibitem [{\citenamefont {Mak}\ and\ \citenamefont {Shan}(2022)}]{Mak2022}%
  \BibitemOpen
  \bibfield  {author} {\bibinfo {author} {\bibfnamefont {K.~F.}\ \bibnamefont
  {Mak}}\ and\ \bibinfo {author} {\bibfnamefont {J.}~\bibnamefont {Shan}},\
  }\bibfield  {title} {\bibinfo {title} {Semiconductor moir{\'e} materials},\
  }\href@noop {} {\bibfield  {journal} {\bibinfo  {journal} {Nature
  Nanotechnology}\ }\textbf {\bibinfo {volume} {17}},\ \bibinfo {pages} {686}
  (\bibinfo {year} {2022})}\BibitemShut {NoStop}%
\bibitem [{\citenamefont {Pixley}\ and\ \citenamefont
  {Volkov}(2025)}]{Pixley2025}%
  \BibitemOpen
  \bibfield  {author} {\bibinfo {author} {\bibfnamefont {J.}~\bibnamefont
  {Pixley}}\ and\ \bibinfo {author} {\bibfnamefont {P.~A.}\ \bibnamefont
  {Volkov}},\ }\bibfield  {title} {\bibinfo {title} {Twisted nodal
  superconductors},\ }\href@noop {} {\bibfield  {journal} {\bibinfo  {journal}
  {arXiv preprint arXiv:2503.23683}\ } (\bibinfo {year} {2025})}\BibitemShut
  {NoStop}%
\bibitem [{\citenamefont {Bernevig}\ \emph {et~al.}(2025)\citenamefont
  {Bernevig}, \citenamefont {Fu}, \citenamefont {Ju}, \citenamefont
  {MacDonald}, \citenamefont {Mak},\ and\ \citenamefont {Shan}}]{Bernevig2025}%
  \BibitemOpen
  \bibfield  {author} {\bibinfo {author} {\bibfnamefont {B.}~\bibnamefont
  {Bernevig}}, \bibinfo {author} {\bibfnamefont {L.}~\bibnamefont {Fu}},
  \bibinfo {author} {\bibfnamefont {L.}~\bibnamefont {Ju}}, \bibinfo {author}
  {\bibfnamefont {A.}~\bibnamefont {MacDonald}}, \bibinfo {author}
  {\bibfnamefont {K.}~\bibnamefont {Mak}},\ and\ \bibinfo {author}
  {\bibfnamefont {J.}~\bibnamefont {Shan}},\ }\bibfield  {title} {\bibinfo
  {title} {Fractional quantization in insulators from hall to chern},\
  }\href@noop {} {\bibfield  {journal} {\bibinfo  {journal} {Nature Physics}\
  ,\ \bibinfo {pages} {1}} (\bibinfo {year} {2025})}\BibitemShut {NoStop}%
\bibitem [{\citenamefont {Cao}\ \emph {et~al.}(2018{\natexlab{a}})\citenamefont
  {Cao}, \citenamefont {Fatemi}, \citenamefont {Demir}, \citenamefont {Fang},
  \citenamefont {Tomarken}, \citenamefont {Luo}, \citenamefont
  {Sanchez-Yamagishi}, \citenamefont {Watanabe}, \citenamefont {Taniguchi},
  \citenamefont {Kaxiras} \emph {et~al.}}]{Cao2018b}%
  \BibitemOpen
  \bibfield  {author} {\bibinfo {author} {\bibfnamefont {Y.}~\bibnamefont
  {Cao}}, \bibinfo {author} {\bibfnamefont {V.}~\bibnamefont {Fatemi}},
  \bibinfo {author} {\bibfnamefont {A.}~\bibnamefont {Demir}}, \bibinfo
  {author} {\bibfnamefont {S.}~\bibnamefont {Fang}}, \bibinfo {author}
  {\bibfnamefont {S.~L.}\ \bibnamefont {Tomarken}}, \bibinfo {author}
  {\bibfnamefont {J.~Y.}\ \bibnamefont {Luo}}, \bibinfo {author} {\bibfnamefont
  {J.~D.}\ \bibnamefont {Sanchez-Yamagishi}}, \bibinfo {author} {\bibfnamefont
  {K.}~\bibnamefont {Watanabe}}, \bibinfo {author} {\bibfnamefont
  {T.}~\bibnamefont {Taniguchi}}, \bibinfo {author} {\bibfnamefont
  {E.}~\bibnamefont {Kaxiras}}, \emph {et~al.},\ }\bibfield  {title} {\bibinfo
  {title} {Correlated insulator behaviour at half-filling in magic-angle
  graphene superlattices},\ }\href@noop {} {\bibfield  {journal} {\bibinfo
  {journal} {Nature}\ }\textbf {\bibinfo {volume} {556}},\ \bibinfo {pages}
  {80} (\bibinfo {year} {2018}{\natexlab{a}})}\BibitemShut {NoStop}%
\bibitem [{\citenamefont {Cao}\ \emph {et~al.}(2018{\natexlab{b}})\citenamefont
  {Cao}, \citenamefont {Fatemi}, \citenamefont {Fang}, \citenamefont
  {Watanabe}, \citenamefont {Taniguchi}, \citenamefont {Kaxiras},\ and\
  \citenamefont {Jarillo-Herrero}}]{Cao2018}%
  \BibitemOpen
  \bibfield  {author} {\bibinfo {author} {\bibfnamefont {Y.}~\bibnamefont
  {Cao}}, \bibinfo {author} {\bibfnamefont {V.}~\bibnamefont {Fatemi}},
  \bibinfo {author} {\bibfnamefont {S.}~\bibnamefont {Fang}}, \bibinfo {author}
  {\bibfnamefont {K.}~\bibnamefont {Watanabe}}, \bibinfo {author}
  {\bibfnamefont {T.}~\bibnamefont {Taniguchi}}, \bibinfo {author}
  {\bibfnamefont {E.}~\bibnamefont {Kaxiras}},\ and\ \bibinfo {author}
  {\bibfnamefont {P.}~\bibnamefont {Jarillo-Herrero}},\ }\bibfield  {title}
  {\bibinfo {title} {Unconventional superconductivity in magic-angle graphene
  superlattices},\ }\href@noop {} {\bibfield  {journal} {\bibinfo  {journal}
  {Nature}\ }\textbf {\bibinfo {volume} {556}},\ \bibinfo {pages} {43}
  (\bibinfo {year} {2018}{\natexlab{b}})}\BibitemShut {NoStop}%
\bibitem [{\citenamefont {Li}\ \emph {et~al.}(2010)\citenamefont {Li},
  \citenamefont {Luican}, \citenamefont {Lopes~dos Santos}, \citenamefont
  {Castro~Neto}, \citenamefont {Reina}, \citenamefont {Kong},\ and\
  \citenamefont {Andrei}}]{Li2010}%
  \BibitemOpen
  \bibfield  {author} {\bibinfo {author} {\bibfnamefont {G.}~\bibnamefont
  {Li}}, \bibinfo {author} {\bibfnamefont {A.}~\bibnamefont {Luican}}, \bibinfo
  {author} {\bibfnamefont {J.}~\bibnamefont {Lopes~dos Santos}}, \bibinfo
  {author} {\bibfnamefont {A.}~\bibnamefont {Castro~Neto}}, \bibinfo {author}
  {\bibfnamefont {A.}~\bibnamefont {Reina}}, \bibinfo {author} {\bibfnamefont
  {J.}~\bibnamefont {Kong}},\ and\ \bibinfo {author} {\bibfnamefont
  {E.}~\bibnamefont {Andrei}},\ }\bibfield  {title} {\bibinfo {title}
  {Observation of van hove singularities in twisted graphene layers},\
  }\href@noop {} {\bibfield  {journal} {\bibinfo  {journal} {Nature physics}\
  }\textbf {\bibinfo {volume} {6}},\ \bibinfo {pages} {109} (\bibinfo {year}
  {2010})}\BibitemShut {NoStop}%
\bibitem [{\citenamefont {Ghiotto}\ \emph {et~al.}(2021)\citenamefont
  {Ghiotto}, \citenamefont {Shih}, \citenamefont {Pereira}, \citenamefont
  {Rhodes}, \citenamefont {Kim}, \citenamefont {Zang}, \citenamefont {Millis},
  \citenamefont {Watanabe}, \citenamefont {Taniguchi}, \citenamefont {Hone}
  \emph {et~al.}}]{Ghiotto2021}%
  \BibitemOpen
  \bibfield  {author} {\bibinfo {author} {\bibfnamefont {A.}~\bibnamefont
  {Ghiotto}}, \bibinfo {author} {\bibfnamefont {E.-M.}\ \bibnamefont {Shih}},
  \bibinfo {author} {\bibfnamefont {G.~S.}\ \bibnamefont {Pereira}}, \bibinfo
  {author} {\bibfnamefont {D.~A.}\ \bibnamefont {Rhodes}}, \bibinfo {author}
  {\bibfnamefont {B.}~\bibnamefont {Kim}}, \bibinfo {author} {\bibfnamefont
  {J.}~\bibnamefont {Zang}}, \bibinfo {author} {\bibfnamefont {A.~J.}\
  \bibnamefont {Millis}}, \bibinfo {author} {\bibfnamefont {K.}~\bibnamefont
  {Watanabe}}, \bibinfo {author} {\bibfnamefont {T.}~\bibnamefont {Taniguchi}},
  \bibinfo {author} {\bibfnamefont {J.~C.}\ \bibnamefont {Hone}}, \emph
  {et~al.},\ }\bibfield  {title} {\bibinfo {title} {Quantum criticality in
  twisted transition metal dichalcogenides},\ }\href@noop {} {\bibfield
  {journal} {\bibinfo  {journal} {Nature}\ }\textbf {\bibinfo {volume} {597}},\
  \bibinfo {pages} {345} (\bibinfo {year} {2021})}\BibitemShut {NoStop}%
\bibitem [{\citenamefont {Zhao}\ \emph
  {et~al.}(2023{\natexlab{a}})\citenamefont {Zhao}, \citenamefont {Shen},
  \citenamefont {Tao}, \citenamefont {Han}, \citenamefont {Kang}, \citenamefont
  {Watanabe}, \citenamefont {Taniguchi}, \citenamefont {Mak},\ and\
  \citenamefont {Shan}}]{Zhao2023}%
  \BibitemOpen
  \bibfield  {author} {\bibinfo {author} {\bibfnamefont {W.}~\bibnamefont
  {Zhao}}, \bibinfo {author} {\bibfnamefont {B.}~\bibnamefont {Shen}}, \bibinfo
  {author} {\bibfnamefont {Z.}~\bibnamefont {Tao}}, \bibinfo {author}
  {\bibfnamefont {Z.}~\bibnamefont {Han}}, \bibinfo {author} {\bibfnamefont
  {K.}~\bibnamefont {Kang}}, \bibinfo {author} {\bibfnamefont {K.}~\bibnamefont
  {Watanabe}}, \bibinfo {author} {\bibfnamefont {T.}~\bibnamefont {Taniguchi}},
  \bibinfo {author} {\bibfnamefont {K.~F.}\ \bibnamefont {Mak}},\ and\ \bibinfo
  {author} {\bibfnamefont {J.}~\bibnamefont {Shan}},\ }\bibfield  {title}
  {\bibinfo {title} {Gate-tunable heavy fermions in a moir{\'e} kondo
  lattice},\ }\href@noop {} {\bibfield  {journal} {\bibinfo  {journal}
  {Nature}\ }\textbf {\bibinfo {volume} {616}},\ \bibinfo {pages} {61}
  (\bibinfo {year} {2023}{\natexlab{a}})}\BibitemShut {NoStop}%
\bibitem [{\citenamefont {Zhao}\ \emph
  {et~al.}(2023{\natexlab{b}})\citenamefont {Zhao}, \citenamefont {Cui},
  \citenamefont {Volkov}, \citenamefont {Yoo}, \citenamefont {Lee},
  \citenamefont {Gardener}, \citenamefont {Akey}, \citenamefont {Engelke},
  \citenamefont {Ronen}, \citenamefont {Zhong} \emph {et~al.}}]{ZhaoS2023}%
  \BibitemOpen
  \bibfield  {author} {\bibinfo {author} {\bibfnamefont {S.~F.}\ \bibnamefont
  {Zhao}}, \bibinfo {author} {\bibfnamefont {X.}~\bibnamefont {Cui}}, \bibinfo
  {author} {\bibfnamefont {P.~A.}\ \bibnamefont {Volkov}}, \bibinfo {author}
  {\bibfnamefont {H.}~\bibnamefont {Yoo}}, \bibinfo {author} {\bibfnamefont
  {S.}~\bibnamefont {Lee}}, \bibinfo {author} {\bibfnamefont {J.~A.}\
  \bibnamefont {Gardener}}, \bibinfo {author} {\bibfnamefont {A.~J.}\
  \bibnamefont {Akey}}, \bibinfo {author} {\bibfnamefont {R.}~\bibnamefont
  {Engelke}}, \bibinfo {author} {\bibfnamefont {Y.}~\bibnamefont {Ronen}},
  \bibinfo {author} {\bibfnamefont {R.}~\bibnamefont {Zhong}}, \emph {et~al.},\
  }\bibfield  {title} {\bibinfo {title} {Time-reversal symmetry breaking
  superconductivity between twisted cuprate superconductors},\ }\href@noop {}
  {\bibfield  {journal} {\bibinfo  {journal} {Science}\ }\textbf {\bibinfo
  {volume} {382}},\ \bibinfo {pages} {1422} (\bibinfo {year}
  {2023}{\natexlab{b}})}\BibitemShut {NoStop}%
\bibitem [{\citenamefont {Castro~Neto}\ \emph {et~al.}(2009)\citenamefont
  {Castro~Neto}, \citenamefont {Guinea}, \citenamefont {Peres}, \citenamefont
  {Novoselov},\ and\ \citenamefont {Geim}}]{CastroNeto2009}%
  \BibitemOpen
  \bibfield  {author} {\bibinfo {author} {\bibfnamefont {A.~H.}\ \bibnamefont
  {Castro~Neto}}, \bibinfo {author} {\bibfnamefont {F.}~\bibnamefont {Guinea}},
  \bibinfo {author} {\bibfnamefont {N.~M.~R.}\ \bibnamefont {Peres}}, \bibinfo
  {author} {\bibfnamefont {K.~S.}\ \bibnamefont {Novoselov}},\ and\ \bibinfo
  {author} {\bibfnamefont {A.~K.}\ \bibnamefont {Geim}},\ }\bibfield  {title}
  {\bibinfo {title} {The electronic properties of graphene},\ }\href
  {https://doi.org/10.1103/RevModPhys.81.109} {\bibfield  {journal} {\bibinfo
  {journal} {Rev. Mod. Phys.}\ }\textbf {\bibinfo {volume} {81}},\ \bibinfo
  {pages} {109} (\bibinfo {year} {2009})}\BibitemShut {NoStop}%
\bibitem [{\citenamefont {Lu}\ \emph {et~al.}(2024)\citenamefont {Lu},
  \citenamefont {Han}, \citenamefont {Yao}, \citenamefont {Reddy},
  \citenamefont {Yang}, \citenamefont {Seo}, \citenamefont {Watanabe},
  \citenamefont {Taniguchi}, \citenamefont {Fu},\ and\ \citenamefont
  {Ju}}]{Lu2024}%
  \BibitemOpen
  \bibfield  {author} {\bibinfo {author} {\bibfnamefont {Z.}~\bibnamefont
  {Lu}}, \bibinfo {author} {\bibfnamefont {T.}~\bibnamefont {Han}}, \bibinfo
  {author} {\bibfnamefont {Y.}~\bibnamefont {Yao}}, \bibinfo {author}
  {\bibfnamefont {A.~P.}\ \bibnamefont {Reddy}}, \bibinfo {author}
  {\bibfnamefont {J.}~\bibnamefont {Yang}}, \bibinfo {author} {\bibfnamefont
  {J.}~\bibnamefont {Seo}}, \bibinfo {author} {\bibfnamefont {K.}~\bibnamefont
  {Watanabe}}, \bibinfo {author} {\bibfnamefont {T.}~\bibnamefont {Taniguchi}},
  \bibinfo {author} {\bibfnamefont {L.}~\bibnamefont {Fu}},\ and\ \bibinfo
  {author} {\bibfnamefont {L.}~\bibnamefont {Ju}},\ }\bibfield  {title}
  {\bibinfo {title} {Fractional quantum anomalous hall effect in multilayer
  graphene},\ }\href@noop {} {\bibfield  {journal} {\bibinfo  {journal}
  {Nature}\ }\textbf {\bibinfo {volume} {626}},\ \bibinfo {pages} {759}
  (\bibinfo {year} {2024})}\BibitemShut {NoStop}%
\bibitem [{\citenamefont {Han}\ \emph {et~al.}(2025)\citenamefont {Han},
  \citenamefont {Lu}, \citenamefont {Hadjri}, \citenamefont {Shi},
  \citenamefont {Wu}, \citenamefont {Xu}, \citenamefont {Yao}, \citenamefont
  {Cotten}, \citenamefont {Sharifi~Sedeh}, \citenamefont {Weldeyesus} \emph
  {et~al.}}]{Han2025}%
  \BibitemOpen
  \bibfield  {author} {\bibinfo {author} {\bibfnamefont {T.}~\bibnamefont
  {Han}}, \bibinfo {author} {\bibfnamefont {Z.}~\bibnamefont {Lu}}, \bibinfo
  {author} {\bibfnamefont {Z.}~\bibnamefont {Hadjri}}, \bibinfo {author}
  {\bibfnamefont {L.}~\bibnamefont {Shi}}, \bibinfo {author} {\bibfnamefont
  {Z.}~\bibnamefont {Wu}}, \bibinfo {author} {\bibfnamefont {W.}~\bibnamefont
  {Xu}}, \bibinfo {author} {\bibfnamefont {Y.}~\bibnamefont {Yao}}, \bibinfo
  {author} {\bibfnamefont {A.~A.}\ \bibnamefont {Cotten}}, \bibinfo {author}
  {\bibfnamefont {O.}~\bibnamefont {Sharifi~Sedeh}}, \bibinfo {author}
  {\bibfnamefont {H.}~\bibnamefont {Weldeyesus}}, \emph {et~al.},\ }\bibfield
  {title} {\bibinfo {title} {Signatures of chiral superconductivity in
  rhombohedral graphene},\ }\href@noop {} {\bibfield  {journal} {\bibinfo
  {journal} {Nature}\ }\textbf {\bibinfo {volume} {643}},\ \bibinfo {pages}
  {654} (\bibinfo {year} {2025})}\BibitemShut {NoStop}%
\bibitem [{\citenamefont {Mounet}\ \emph
  {et~al.}(2018{\natexlab{a}})\citenamefont {Mounet}, \citenamefont
  {Gibertini}, \citenamefont {Schwaller}, \citenamefont {Campi}, \citenamefont
  {Merkys}, \citenamefont {Marrazzo}, \citenamefont {Sohier}, \citenamefont
  {Castelli}, \citenamefont {Cepellotti}, \citenamefont {Pizzi} \emph
  {et~al.}}]{Mounet2018}%
  \BibitemOpen
  \bibfield  {author} {\bibinfo {author} {\bibfnamefont {N.}~\bibnamefont
  {Mounet}}, \bibinfo {author} {\bibfnamefont {M.}~\bibnamefont {Gibertini}},
  \bibinfo {author} {\bibfnamefont {P.}~\bibnamefont {Schwaller}}, \bibinfo
  {author} {\bibfnamefont {D.}~\bibnamefont {Campi}}, \bibinfo {author}
  {\bibfnamefont {A.}~\bibnamefont {Merkys}}, \bibinfo {author} {\bibfnamefont
  {A.}~\bibnamefont {Marrazzo}}, \bibinfo {author} {\bibfnamefont
  {T.}~\bibnamefont {Sohier}}, \bibinfo {author} {\bibfnamefont {I.~E.}\
  \bibnamefont {Castelli}}, \bibinfo {author} {\bibfnamefont {A.}~\bibnamefont
  {Cepellotti}}, \bibinfo {author} {\bibfnamefont {G.}~\bibnamefont {Pizzi}},
  \emph {et~al.},\ }\bibfield  {title} {\bibinfo {title} {Two-dimensional
  materials from high-throughput computational exfoliation of experimentally
  known compounds},\ }\href@noop {} {\bibfield  {journal} {\bibinfo  {journal}
  {Nature nanotechnology}\ }\textbf {\bibinfo {volume} {13}},\ \bibinfo {pages}
  {246} (\bibinfo {year} {2018}{\natexlab{a}})}\BibitemShut {NoStop}%
\bibitem [{\citenamefont {Campi}\ \emph {et~al.}(2023)\citenamefont {Campi},
  \citenamefont {Mounet}, \citenamefont {Gibertini}, \citenamefont {Pizzi},\
  and\ \citenamefont {Marzari}}]{campi2023expansion}%
  \BibitemOpen
  \bibfield  {author} {\bibinfo {author} {\bibfnamefont {D.}~\bibnamefont
  {Campi}}, \bibinfo {author} {\bibfnamefont {N.}~\bibnamefont {Mounet}},
  \bibinfo {author} {\bibfnamefont {M.}~\bibnamefont {Gibertini}}, \bibinfo
  {author} {\bibfnamefont {G.}~\bibnamefont {Pizzi}},\ and\ \bibinfo {author}
  {\bibfnamefont {N.}~\bibnamefont {Marzari}},\ }\bibfield  {title} {\bibinfo
  {title} {Expansion of the materials cloud 2d database},\ }\href
  {https://doi.org/https://doi.org/10.1021/acsnano.2c11510} {\bibfield
  {journal} {\bibinfo  {journal} {ACS nano}\ }\textbf {\bibinfo {volume}
  {17}},\ \bibinfo {pages} {11268} (\bibinfo {year} {2023})}\BibitemShut
  {NoStop}%
\bibitem [{\citenamefont {Mounet}\ \emph
  {et~al.}(2018{\natexlab{b}})\citenamefont {Mounet}, \citenamefont
  {Gibertini}, \citenamefont {Schwaller}, \citenamefont {Campi}, \citenamefont
  {Merkys}, \citenamefont {Marrazzo}, \citenamefont {Sohier}, \citenamefont
  {Castelli}, \citenamefont {Cepellotti}, \citenamefont {Pizzi} \emph
  {et~al.}}]{mounet2018two}%
  \BibitemOpen
  \bibfield  {author} {\bibinfo {author} {\bibfnamefont {N.}~\bibnamefont
  {Mounet}}, \bibinfo {author} {\bibfnamefont {M.}~\bibnamefont {Gibertini}},
  \bibinfo {author} {\bibfnamefont {P.}~\bibnamefont {Schwaller}}, \bibinfo
  {author} {\bibfnamefont {D.}~\bibnamefont {Campi}}, \bibinfo {author}
  {\bibfnamefont {A.}~\bibnamefont {Merkys}}, \bibinfo {author} {\bibfnamefont
  {A.}~\bibnamefont {Marrazzo}}, \bibinfo {author} {\bibfnamefont
  {T.}~\bibnamefont {Sohier}}, \bibinfo {author} {\bibfnamefont {I.~E.}\
  \bibnamefont {Castelli}}, \bibinfo {author} {\bibfnamefont {A.}~\bibnamefont
  {Cepellotti}}, \bibinfo {author} {\bibfnamefont {G.}~\bibnamefont {Pizzi}},
  \emph {et~al.},\ }\bibfield  {title} {\bibinfo {title} {Two-dimensional
  materials from high-throughput computational exfoliation of experimentally
  known compounds},\ }\href
  {https://doi.org/https://doi.org/10.1038/s41565-017-0035-5} {\bibfield
  {journal} {\bibinfo  {journal} {Nature nanotechnology}\ }\textbf {\bibinfo
  {volume} {13}},\ \bibinfo {pages} {246} (\bibinfo {year}
  {2018}{\natexlab{b}})}\BibitemShut {NoStop}%
\bibitem [{\citenamefont {Jiang}\ \emph {et~al.}(2024)\citenamefont {Jiang},
  \citenamefont {Petralanda}, \citenamefont {Skorupskii}, \citenamefont {Xu},
  \citenamefont {Pi}, \citenamefont {C{\u{a}}lug{\u{a}}ru}, \citenamefont {Hu},
  \citenamefont {Xie}, \citenamefont {Mustaf}, \citenamefont {H{\"o}hn} \emph
  {et~al.}}]{jiang20242d}%
  \BibitemOpen
  \bibfield  {author} {\bibinfo {author} {\bibfnamefont {Y.}~\bibnamefont
  {Jiang}}, \bibinfo {author} {\bibfnamefont {U.}~\bibnamefont {Petralanda}},
  \bibinfo {author} {\bibfnamefont {G.}~\bibnamefont {Skorupskii}}, \bibinfo
  {author} {\bibfnamefont {Q.}~\bibnamefont {Xu}}, \bibinfo {author}
  {\bibfnamefont {H.}~\bibnamefont {Pi}}, \bibinfo {author} {\bibfnamefont
  {D.}~\bibnamefont {C{\u{a}}lug{\u{a}}ru}}, \bibinfo {author} {\bibfnamefont
  {H.}~\bibnamefont {Hu}}, \bibinfo {author} {\bibfnamefont {J.}~\bibnamefont
  {Xie}}, \bibinfo {author} {\bibfnamefont {R.~A.}\ \bibnamefont {Mustaf}},
  \bibinfo {author} {\bibfnamefont {P.}~\bibnamefont {H{\"o}hn}}, \emph
  {et~al.},\ }\bibfield  {title} {\bibinfo {title} {2d theoretically twistable
  material database},\ }\href@noop {} {\bibfield  {journal} {\bibinfo
  {journal} {arXiv preprint arXiv:2411.09741}\ } (\bibinfo {year}
  {2024})}\BibitemShut {NoStop}%
\bibitem [{\citenamefont {Cr\'epel}\ and\ \citenamefont
  {Cano}(2025)}]{Crepel2025}%
  \BibitemOpen
  \bibfield  {author} {\bibinfo {author} {\bibfnamefont {V.}~\bibnamefont
  {Cr\'epel}}\ and\ \bibinfo {author} {\bibfnamefont {J.}~\bibnamefont
  {Cano}},\ }\bibfield  {title} {\bibinfo {title} {Efficient prediction of
  superlattice and anomalous miniband topology from quantum geometry},\ }\href
  {https://doi.org/10.1103/PhysRevX.15.011004} {\bibfield  {journal} {\bibinfo
  {journal} {Phys. Rev. X}\ }\textbf {\bibinfo {volume} {15}},\ \bibinfo
  {pages} {011004} (\bibinfo {year} {2025})}\BibitemShut {NoStop}%
\bibitem [{\citenamefont {Lhachemi}\ \emph {et~al.}(2025)\citenamefont
  {Lhachemi}, \citenamefont {Cr{\'e}pel},\ and\ \citenamefont
  {Cano}}]{Lhachemi2025}%
  \BibitemOpen
  \bibfield  {author} {\bibinfo {author} {\bibfnamefont {M.~N.~Y.}\
  \bibnamefont {Lhachemi}}, \bibinfo {author} {\bibfnamefont {V.}~\bibnamefont
  {Cr{\'e}pel}},\ and\ \bibinfo {author} {\bibfnamefont {J.}~\bibnamefont
  {Cano}},\ }\bibfield  {title} {\bibinfo {title} {Efficient prediction of
  topological superlattice bands with spin-orbit coupling},\ }\href@noop {}
  {\bibfield  {journal} {\bibinfo  {journal} {arXiv preprint arXiv:2511.17483}\
  } (\bibinfo {year} {2025})}\BibitemShut {NoStop}%
\bibitem [{\citenamefont {Nakatsuji}\ \emph {et~al.}()\citenamefont
  {Nakatsuji}, \citenamefont {Cano},\ and\ \citenamefont
  {Cr\'epel}}]{Nakatsuji2025}%
  \BibitemOpen
  \bibfield  {author} {\bibinfo {author} {\bibfnamefont {N.}~\bibnamefont
  {Nakatsuji}}, \bibinfo {author} {\bibfnamefont {J.}~\bibnamefont {Cano}},\
  and\ \bibinfo {author} {\bibfnamefont {V.}~\bibnamefont {Cr\'epel}},\
  }\href@noop {} {\bibinfo {title} {High-throughput discovery of moir\'e
  homobilayers guided by topology and energetics}},\ \bibinfo {note} {to
  appear}\BibitemShut {NoStop}%
\bibitem [{\citenamefont {Zhang}\ \emph {et~al.}(2022)\citenamefont {Zhang},
  \citenamefont {Regnault}, \citenamefont {Bernevig}, \citenamefont {Dai},\
  and\ \citenamefont {Weng}}]{Zhang2022}%
  \BibitemOpen
  \bibfield  {author} {\bibinfo {author} {\bibfnamefont {T.}~\bibnamefont
  {Zhang}}, \bibinfo {author} {\bibfnamefont {N.}~\bibnamefont {Regnault}},
  \bibinfo {author} {\bibfnamefont {B.~A.}\ \bibnamefont {Bernevig}}, \bibinfo
  {author} {\bibfnamefont {X.}~\bibnamefont {Dai}},\ and\ \bibinfo {author}
  {\bibfnamefont {H.}~\bibnamefont {Weng}},\ }\bibfield  {title} {\bibinfo
  {title} {$o(n$) ab initio calculation scheme for large-scale moir\'e
  structures},\ }\href {https://doi.org/10.1103/PhysRevB.105.125127} {\bibfield
   {journal} {\bibinfo  {journal} {Phys. Rev. B}\ }\textbf {\bibinfo {volume}
  {105}},\ \bibinfo {pages} {125127} (\bibinfo {year} {2022})}\BibitemShut
  {NoStop}%
\bibitem [{\citenamefont {Fang}\ and\ \citenamefont
  {Kaxiras}(2016)}]{Fang2016}%
  \BibitemOpen
  \bibfield  {author} {\bibinfo {author} {\bibfnamefont {S.}~\bibnamefont
  {Fang}}\ and\ \bibinfo {author} {\bibfnamefont {E.}~\bibnamefont {Kaxiras}},\
  }\bibfield  {title} {\bibinfo {title} {Electronic structure theory of weakly
  interacting bilayers},\ }\href {https://doi.org/10.1103/PhysRevB.93.235153}
  {\bibfield  {journal} {\bibinfo  {journal} {Phys. Rev. B}\ }\textbf {\bibinfo
  {volume} {93}},\ \bibinfo {pages} {235153} (\bibinfo {year}
  {2016})}\BibitemShut {NoStop}%
\bibitem [{\citenamefont {Massatt}\ \emph {et~al.}(2017)\citenamefont
  {Massatt}, \citenamefont {Luskin},\ and\ \citenamefont
  {Ortner}}]{massatt2017}%
  \BibitemOpen
  \bibfield  {author} {\bibinfo {author} {\bibfnamefont {D.}~\bibnamefont
  {Massatt}}, \bibinfo {author} {\bibfnamefont {M.}~\bibnamefont {Luskin}},\
  and\ \bibinfo {author} {\bibfnamefont {C.}~\bibnamefont {Ortner}},\
  }\bibfield  {title} {\bibinfo {title} {Electronic density of states for
  incommensurate layers},\ }\href@noop {} {\bibfield  {journal} {\bibinfo
  {journal} {Multiscale Modeling \& Simulation}\ }\textbf {\bibinfo {volume}
  {15}},\ \bibinfo {pages} {476} (\bibinfo {year} {2017})}\BibitemShut
  {NoStop}%
\bibitem [{\citenamefont {Carr}\ \emph {et~al.}(2017)\citenamefont {Carr},
  \citenamefont {Massatt}, \citenamefont {Fang}, \citenamefont {Cazeaux},
  \citenamefont {Luskin},\ and\ \citenamefont {Kaxiras}}]{Carr2017}%
  \BibitemOpen
  \bibfield  {author} {\bibinfo {author} {\bibfnamefont {S.}~\bibnamefont
  {Carr}}, \bibinfo {author} {\bibfnamefont {D.}~\bibnamefont {Massatt}},
  \bibinfo {author} {\bibfnamefont {S.}~\bibnamefont {Fang}}, \bibinfo {author}
  {\bibfnamefont {P.}~\bibnamefont {Cazeaux}}, \bibinfo {author} {\bibfnamefont
  {M.}~\bibnamefont {Luskin}},\ and\ \bibinfo {author} {\bibfnamefont
  {E.}~\bibnamefont {Kaxiras}},\ }\bibfield  {title} {\bibinfo {title}
  {Twistronics: Manipulating the electronic properties of two-dimensional
  layered structures through their twist angle},\ }\href
  {https://doi.org/10.1103/PhysRevB.95.075420} {\bibfield  {journal} {\bibinfo
  {journal} {Phys. Rev. B}\ }\textbf {\bibinfo {volume} {95}},\ \bibinfo
  {pages} {075420} (\bibinfo {year} {2017})}\BibitemShut {NoStop}%
\bibitem [{\citenamefont {Cances}\ \emph {et~al.}(2017)\citenamefont {Cances},
  \citenamefont {Cazeaux},\ and\ \citenamefont {Luskin}}]{Cances2017}%
  \BibitemOpen
  \bibfield  {author} {\bibinfo {author} {\bibfnamefont {E.}~\bibnamefont
  {Cances}}, \bibinfo {author} {\bibfnamefont {P.}~\bibnamefont {Cazeaux}},\
  and\ \bibinfo {author} {\bibfnamefont {M.}~\bibnamefont {Luskin}},\
  }\bibfield  {title} {\bibinfo {title} {Generalized kubo formulas for the
  transport properties of incommensurate 2d atomic heterostructures},\
  }\href@noop {} {\bibfield  {journal} {\bibinfo  {journal} {Journal of
  Mathematical Physics}\ }\textbf {\bibinfo {volume} {58}} (\bibinfo {year}
  {2017})}\BibitemShut {NoStop}%
\bibitem [{\citenamefont {Song}\ \emph {et~al.}(2019)\citenamefont {Song},
  \citenamefont {Wang}, \citenamefont {Shi}, \citenamefont {Li}, \citenamefont
  {Fang},\ and\ \citenamefont {Bernevig}}]{Song2019}%
  \BibitemOpen
  \bibfield  {author} {\bibinfo {author} {\bibfnamefont {Z.}~\bibnamefont
  {Song}}, \bibinfo {author} {\bibfnamefont {Z.}~\bibnamefont {Wang}}, \bibinfo
  {author} {\bibfnamefont {W.}~\bibnamefont {Shi}}, \bibinfo {author}
  {\bibfnamefont {G.}~\bibnamefont {Li}}, \bibinfo {author} {\bibfnamefont
  {C.}~\bibnamefont {Fang}},\ and\ \bibinfo {author} {\bibfnamefont {B.~A.}\
  \bibnamefont {Bernevig}},\ }\bibfield  {title} {\bibinfo {title} {All magic
  angles in twisted bilayer graphene are topological},\ }\href
  {https://doi.org/10.1103/PhysRevLett.123.036401} {\bibfield  {journal}
  {\bibinfo  {journal} {Phys. Rev. Lett.}\ }\textbf {\bibinfo {volume} {123}},\
  \bibinfo {pages} {036401} (\bibinfo {year} {2019})}\BibitemShut {NoStop}%
\bibitem [{\citenamefont {Devakul}\ \emph {et~al.}(2021)\citenamefont
  {Devakul}, \citenamefont {Cr{\'e}pel}, \citenamefont {Zhang},\ and\
  \citenamefont {Fu}}]{devakul2021magic}%
  \BibitemOpen
  \bibfield  {author} {\bibinfo {author} {\bibfnamefont {T.}~\bibnamefont
  {Devakul}}, \bibinfo {author} {\bibfnamefont {V.}~\bibnamefont {Cr{\'e}pel}},
  \bibinfo {author} {\bibfnamefont {Y.}~\bibnamefont {Zhang}},\ and\ \bibinfo
  {author} {\bibfnamefont {L.}~\bibnamefont {Fu}},\ }\bibfield  {title}
  {\bibinfo {title} {Magic in twisted transition metal dichalcogenide
  bilayers},\ }\href
  {https://doi.org/https://doi.org/10.1038/s41467-021-27042-9} {\bibfield
  {journal} {\bibinfo  {journal} {Nat. Commun.}\ }\textbf {\bibinfo {volume}
  {12}},\ \bibinfo {pages} {6730} (\bibinfo {year} {2021})}\BibitemShut
  {NoStop}%
\bibitem [{\citenamefont {Zhong}\ \emph {et~al.}(2023)\citenamefont {Zhong},
  \citenamefont {Yu}, \citenamefont {Su}, \citenamefont {Gong},\ and\
  \citenamefont {Xiang}}]{zhong2023transferable}%
  \BibitemOpen
  \bibfield  {author} {\bibinfo {author} {\bibfnamefont {Y.}~\bibnamefont
  {Zhong}}, \bibinfo {author} {\bibfnamefont {H.}~\bibnamefont {Yu}}, \bibinfo
  {author} {\bibfnamefont {M.}~\bibnamefont {Su}}, \bibinfo {author}
  {\bibfnamefont {X.}~\bibnamefont {Gong}},\ and\ \bibinfo {author}
  {\bibfnamefont {H.}~\bibnamefont {Xiang}},\ }\bibfield  {title} {\bibinfo
  {title} {Transferable equivariant graph neural networks for the hamiltonians
  of molecules and solids},\ }\href
  {https://doi.org/https://doi.org/10.1038/s41524-023-01130-4} {\bibfield
  {journal} {\bibinfo  {journal} {npj Comput. Mater.}\ }\textbf {\bibinfo
  {volume} {9}},\ \bibinfo {pages} {182} (\bibinfo {year} {2023})}\BibitemShut
  {NoStop}%
\bibitem [{\citenamefont {Gong}\ \emph {et~al.}(2023)\citenamefont {Gong},
  \citenamefont {Li}, \citenamefont {Zou}, \citenamefont {Xu}, \citenamefont
  {Duan},\ and\ \citenamefont {Xu}}]{gong2023general}%
  \BibitemOpen
  \bibfield  {author} {\bibinfo {author} {\bibfnamefont {X.}~\bibnamefont
  {Gong}}, \bibinfo {author} {\bibfnamefont {H.}~\bibnamefont {Li}}, \bibinfo
  {author} {\bibfnamefont {N.}~\bibnamefont {Zou}}, \bibinfo {author}
  {\bibfnamefont {R.}~\bibnamefont {Xu}}, \bibinfo {author} {\bibfnamefont
  {W.}~\bibnamefont {Duan}},\ and\ \bibinfo {author} {\bibfnamefont
  {Y.}~\bibnamefont {Xu}},\ }\bibfield  {title} {\bibinfo {title} {General
  framework for e (3)-equivariant neural network representation of density
  functional theory hamiltonian},\ }\href
  {https://doi.org/https://doi.org/10.1038/s41467-023-38468-8} {\bibfield
  {journal} {\bibinfo  {journal} {Nat. Commun.}\ }\textbf {\bibinfo {volume}
  {14}},\ \bibinfo {pages} {2848} (\bibinfo {year} {2023})}\BibitemShut
  {NoStop}%
\bibitem [{\citenamefont {Qi}\ \emph {et~al.}(2025)\citenamefont {Qi},
  \citenamefont {Gong},\ and\ \citenamefont {Yan}}]{qi2025bridging}%
  \BibitemOpen
  \bibfield  {author} {\bibinfo {author} {\bibfnamefont {Y.}~\bibnamefont
  {Qi}}, \bibinfo {author} {\bibfnamefont {W.}~\bibnamefont {Gong}},\ and\
  \bibinfo {author} {\bibfnamefont {Q.}~\bibnamefont {Yan}},\ }\bibfield
  {title} {\bibinfo {title} {Bridging deep learning force fields and electronic
  structures with a physics-informed approach},\ }\href
  {https://doi.org/https://doi.org/10.1038/s41524-025-01668-5} {\bibfield
  {journal} {\bibinfo  {journal} {npj Comput. Mater.}\ }\textbf {\bibinfo
  {volume} {11}},\ \bibinfo {pages} {177} (\bibinfo {year} {2025})}\BibitemShut
  {NoStop}%
\bibitem [{\citenamefont {Pizzi}\ \emph {et~al.}(2020)\citenamefont {Pizzi},
  \citenamefont {Vitale}, \citenamefont {Arita}, \citenamefont {Blugel},
  \citenamefont {Freimuth}, \citenamefont {G{\'{e}}ranton}, \citenamefont
  {Gibertini}, \citenamefont {Gresch}, \citenamefont {Johnson}, \citenamefont
  {Koretsune}, \citenamefont {Iba{\~{n}}ez-Azpiroz}, \citenamefont {Lee},
  \citenamefont {Lihm}, \citenamefont {Marchand}, \citenamefont {Marrazzo},
  \citenamefont {Mokrousov}, \citenamefont {Mustafa}, \citenamefont {Nohara},
  \citenamefont {Nomura}, \citenamefont {Paulatto}, \citenamefont
  {Ponc{\'{e}}}, \citenamefont {Ponweiser}, \citenamefont {Qiao}, \citenamefont
  {Thole}, \citenamefont {Tsirkin}, \citenamefont {Wierzbowska}, \citenamefont
  {Marzari}, \citenamefont {Vanderbilt}, \citenamefont {Souza}, \citenamefont
  {Mostofi},\ and\ \citenamefont {Yates}}]{Pizzi2020}%
  \BibitemOpen
  \bibfield  {author} {\bibinfo {author} {\bibfnamefont {G.}~\bibnamefont
  {Pizzi}}, \bibinfo {author} {\bibfnamefont {V.}~\bibnamefont {Vitale}},
  \bibinfo {author} {\bibfnamefont {R.}~\bibnamefont {Arita}}, \bibinfo
  {author} {\bibfnamefont {S.}~\bibnamefont {Blugel}}, \bibinfo {author}
  {\bibfnamefont {F.}~\bibnamefont {Freimuth}}, \bibinfo {author}
  {\bibfnamefont {G.}~\bibnamefont {G{\'{e}}ranton}}, \bibinfo {author}
  {\bibfnamefont {M.}~\bibnamefont {Gibertini}}, \bibinfo {author}
  {\bibfnamefont {D.}~\bibnamefont {Gresch}}, \bibinfo {author} {\bibfnamefont
  {C.}~\bibnamefont {Johnson}}, \bibinfo {author} {\bibfnamefont
  {T.}~\bibnamefont {Koretsune}}, \bibinfo {author} {\bibfnamefont
  {J.}~\bibnamefont {Iba{\~{n}}ez-Azpiroz}}, \bibinfo {author} {\bibfnamefont
  {H.}~\bibnamefont {Lee}}, \bibinfo {author} {\bibfnamefont {J.-M.}\
  \bibnamefont {Lihm}}, \bibinfo {author} {\bibfnamefont {D.}~\bibnamefont
  {Marchand}}, \bibinfo {author} {\bibfnamefont {A.}~\bibnamefont {Marrazzo}},
  \bibinfo {author} {\bibfnamefont {Y.}~\bibnamefont {Mokrousov}}, \bibinfo
  {author} {\bibfnamefont {J.~I.}\ \bibnamefont {Mustafa}}, \bibinfo {author}
  {\bibfnamefont {Y.}~\bibnamefont {Nohara}}, \bibinfo {author} {\bibfnamefont
  {Y.}~\bibnamefont {Nomura}}, \bibinfo {author} {\bibfnamefont
  {L.}~\bibnamefont {Paulatto}}, \bibinfo {author} {\bibfnamefont
  {S.}~\bibnamefont {Ponc{\'{e}}}}, \bibinfo {author} {\bibfnamefont
  {T.}~\bibnamefont {Ponweiser}}, \bibinfo {author} {\bibfnamefont
  {J.}~\bibnamefont {Qiao}}, \bibinfo {author} {\bibfnamefont {F.}~\bibnamefont
  {Thole}}, \bibinfo {author} {\bibfnamefont {S.~S.}\ \bibnamefont {Tsirkin}},
  \bibinfo {author} {\bibfnamefont {M.}~\bibnamefont {Wierzbowska}}, \bibinfo
  {author} {\bibfnamefont {N.}~\bibnamefont {Marzari}}, \bibinfo {author}
  {\bibfnamefont {D.}~\bibnamefont {Vanderbilt}}, \bibinfo {author}
  {\bibfnamefont {I.}~\bibnamefont {Souza}}, \bibinfo {author} {\bibfnamefont
  {A.~A.}\ \bibnamefont {Mostofi}},\ and\ \bibinfo {author} {\bibfnamefont
  {J.~R.}\ \bibnamefont {Yates}},\ }\bibfield  {title} {\bibinfo {title}
  {Wannier90 as a community code: new features and applications},\ }\href
  {https://doi.org/10.1088/1361-648x/ab51ff} {\bibfield  {journal} {\bibinfo
  {journal} {J. Phys. Condens. Matter}\ }\textbf {\bibinfo {volume} {32}},\
  \bibinfo {pages} {165902} (\bibinfo {year} {2020})}\BibitemShut {NoStop}%
\bibitem [{\citenamefont {Carr}\ \emph {et~al.}(2019)\citenamefont {Carr},
  \citenamefont {Fang}, \citenamefont {Zhu},\ and\ \citenamefont
  {Kaxiras}}]{Carr2019}%
  \BibitemOpen
  \bibfield  {author} {\bibinfo {author} {\bibfnamefont {S.}~\bibnamefont
  {Carr}}, \bibinfo {author} {\bibfnamefont {S.}~\bibnamefont {Fang}}, \bibinfo
  {author} {\bibfnamefont {Z.}~\bibnamefont {Zhu}},\ and\ \bibinfo {author}
  {\bibfnamefont {E.}~\bibnamefont {Kaxiras}},\ }\bibfield  {title} {\bibinfo
  {title} {Exact continuum model for low-energy electronic states of twisted
  bilayer graphene},\ }\href {https://doi.org/10.1103/PhysRevResearch.1.013001}
  {\bibfield  {journal} {\bibinfo  {journal} {Phys. Rev. Res.}\ }\textbf
  {\bibinfo {volume} {1}},\ \bibinfo {pages} {013001} (\bibinfo {year}
  {2019})}\BibitemShut {NoStop}%
\bibitem [{\citenamefont {Carr}\ \emph {et~al.}(2020)\citenamefont {Carr},
  \citenamefont {Fang},\ and\ \citenamefont {Kaxiras}}]{carr2020electronic}%
  \BibitemOpen
  \bibfield  {author} {\bibinfo {author} {\bibfnamefont {S.}~\bibnamefont
  {Carr}}, \bibinfo {author} {\bibfnamefont {S.}~\bibnamefont {Fang}},\ and\
  \bibinfo {author} {\bibfnamefont {E.}~\bibnamefont {Kaxiras}},\ }\bibfield
  {title} {\bibinfo {title} {Electronic-structure methods for twisted moir{\'e}
  layers},\ }\href {https://doi.org/https://doi.org/10.1038/s41578-020-0214-0}
  {\bibfield  {journal} {\bibinfo  {journal} {Nat. Rev. Mater.}\ }\textbf
  {\bibinfo {volume} {5}},\ \bibinfo {pages} {748} (\bibinfo {year}
  {2020})}\BibitemShut {NoStop}%
\bibitem [{\citenamefont {Moldovan}\ \emph {et~al.}(2020)\citenamefont
  {Moldovan}, \citenamefont {Anđelković},\ and\ \citenamefont
  {Peeters}}]{dean_moldovan_2020_4010216}%
  \BibitemOpen
  \bibfield  {author} {\bibinfo {author} {\bibfnamefont {D.}~\bibnamefont
  {Moldovan}}, \bibinfo {author} {\bibfnamefont {M.}~\bibnamefont
  {Anđelković}},\ and\ \bibinfo {author} {\bibfnamefont {F.}~\bibnamefont
  {Peeters}},\ }\href {https://doi.org/10.5281/zenodo.4010216} {\bibinfo
  {title} {pybinding v0.9.5: a python package for tight- binding calculations}}
  (\bibinfo {year} {2020})\BibitemShut {NoStop}%
\bibitem [{\citenamefont {Varjas}\ \emph {et~al.}(2020)\citenamefont {Varjas},
  \citenamefont {Fruchart}, \citenamefont {Akhmerov},\ and\ \citenamefont
  {Perez-Piskunow}}]{kpmchern}%
  \BibitemOpen
  \bibfield  {author} {\bibinfo {author} {\bibfnamefont {D.}~\bibnamefont
  {Varjas}}, \bibinfo {author} {\bibfnamefont {M.}~\bibnamefont {Fruchart}},
  \bibinfo {author} {\bibfnamefont {A.~R.}\ \bibnamefont {Akhmerov}},\ and\
  \bibinfo {author} {\bibfnamefont {P.~M.}\ \bibnamefont {Perez-Piskunow}},\
  }\bibfield  {title} {\bibinfo {title} {Computation of topological phase
  diagram of disordered
  ${\mathrm{pb}}_{1\ensuremath{-}x}{\mathrm{sn}}_{x}\mathrm{Te}$ using the
  kernel polynomial method},\ }\href
  {https://doi.org/10.1103/PhysRevResearch.2.013229} {\bibfield  {journal}
  {\bibinfo  {journal} {Phys. Rev. Res.}\ }\textbf {\bibinfo {volume} {2}},\
  \bibinfo {pages} {013229} (\bibinfo {year} {2020})}\BibitemShut {NoStop}%
\bibitem [{\citenamefont {Peng}\ \emph {et~al.}(2016)\citenamefont {Peng},
  \citenamefont {Yang}, \citenamefont {Perdew},\ and\ \citenamefont
  {Sun}}]{peng2016versatile}%
  \BibitemOpen
  \bibfield  {author} {\bibinfo {author} {\bibfnamefont {H.}~\bibnamefont
  {Peng}}, \bibinfo {author} {\bibfnamefont {Z.-H.}\ \bibnamefont {Yang}},
  \bibinfo {author} {\bibfnamefont {J.~P.}\ \bibnamefont {Perdew}},\ and\
  \bibinfo {author} {\bibfnamefont {J.}~\bibnamefont {Sun}},\ }\bibfield
  {title} {\bibinfo {title} {Versatile van der waals density functional based
  on a meta-generalized gradient approximation},\ }\href@noop {} {\bibfield
  {journal} {\bibinfo  {journal} {Physical Review X}\ }\textbf {\bibinfo
  {volume} {6}},\ \bibinfo {pages} {041005} (\bibinfo {year}
  {2016})}\BibitemShut {NoStop}%
\bibitem [{\citenamefont {Zhang}\ \emph {et~al.}(2016)\citenamefont {Zhang},
  \citenamefont {Wang}, \citenamefont {Chen}, \citenamefont {Wang},\ and\
  \citenamefont {Wee}}]{zhang2016van}%
  \BibitemOpen
  \bibfield  {author} {\bibinfo {author} {\bibfnamefont {W.}~\bibnamefont
  {Zhang}}, \bibinfo {author} {\bibfnamefont {Q.}~\bibnamefont {Wang}},
  \bibinfo {author} {\bibfnamefont {Y.}~\bibnamefont {Chen}}, \bibinfo {author}
  {\bibfnamefont {Z.}~\bibnamefont {Wang}},\ and\ \bibinfo {author}
  {\bibfnamefont {A.~T.}\ \bibnamefont {Wee}},\ }\bibfield  {title} {\bibinfo
  {title} {Van der waals stacked 2d layered materials for optoelectronics},\
  }\href@noop {} {\bibfield  {journal} {\bibinfo  {journal} {2D Materials}\
  }\textbf {\bibinfo {volume} {3}},\ \bibinfo {pages} {022001} (\bibinfo {year}
  {2016})}\BibitemShut {NoStop}%
\bibitem [{\citenamefont {Mustonen}\ \emph {et~al.}(2022)\citenamefont
  {Mustonen}, \citenamefont {Hofer}, \citenamefont {Kotrusz}, \citenamefont
  {Markevich}, \citenamefont {Hulman}, \citenamefont {Mangler}, \citenamefont
  {Susi}, \citenamefont {Pennycook}, \citenamefont {Hricovini}, \citenamefont
  {Richter} \emph {et~al.}}]{mustonen2022toward}%
  \BibitemOpen
  \bibfield  {author} {\bibinfo {author} {\bibfnamefont {K.}~\bibnamefont
  {Mustonen}}, \bibinfo {author} {\bibfnamefont {C.}~\bibnamefont {Hofer}},
  \bibinfo {author} {\bibfnamefont {P.}~\bibnamefont {Kotrusz}}, \bibinfo
  {author} {\bibfnamefont {A.}~\bibnamefont {Markevich}}, \bibinfo {author}
  {\bibfnamefont {M.}~\bibnamefont {Hulman}}, \bibinfo {author} {\bibfnamefont
  {C.}~\bibnamefont {Mangler}}, \bibinfo {author} {\bibfnamefont
  {T.}~\bibnamefont {Susi}}, \bibinfo {author} {\bibfnamefont {T.~J.}\
  \bibnamefont {Pennycook}}, \bibinfo {author} {\bibfnamefont {K.}~\bibnamefont
  {Hricovini}}, \bibinfo {author} {\bibfnamefont {C.}~\bibnamefont {Richter}},
  \emph {et~al.},\ }\bibfield  {title} {\bibinfo {title} {Toward exotic layered
  materials: 2d cuprous iodide},\ }\href@noop {} {\bibfield  {journal}
  {\bibinfo  {journal} {Advanced Materials}\ }\textbf {\bibinfo {volume}
  {34}},\ \bibinfo {pages} {2106922} (\bibinfo {year} {2022})}\BibitemShut
  {NoStop}%
\bibitem [{\citenamefont {Beckmann}(2010)}]{beckmann2010review}%
  \BibitemOpen
  \bibfield  {author} {\bibinfo {author} {\bibfnamefont {P.~A.}\ \bibnamefont
  {Beckmann}},\ }\bibfield  {title} {\bibinfo {title} {A review of polytypism
  in lead iodide},\ }\href@noop {} {\bibfield  {journal} {\bibinfo  {journal}
  {Crystal Research and Technology}\ }\textbf {\bibinfo {volume} {45}},\
  \bibinfo {pages} {455} (\bibinfo {year} {2010})}\BibitemShut {NoStop}%
\bibitem [{\citenamefont {Xu}\ \emph {et~al.}(2016)\citenamefont {Xu},
  \citenamefont {Wu}, \citenamefont {Cui}, \citenamefont {Ni}, \citenamefont
  {Xu}, \citenamefont {Cai}, \citenamefont {Hong}, \citenamefont {Fang},
  \citenamefont {Wang}, \citenamefont {Zhu} \emph {et~al.}}]{xu2016formation}%
  \BibitemOpen
  \bibfield  {author} {\bibinfo {author} {\bibfnamefont {H.}~\bibnamefont
  {Xu}}, \bibinfo {author} {\bibfnamefont {Y.}~\bibnamefont {Wu}}, \bibinfo
  {author} {\bibfnamefont {J.}~\bibnamefont {Cui}}, \bibinfo {author}
  {\bibfnamefont {C.}~\bibnamefont {Ni}}, \bibinfo {author} {\bibfnamefont
  {F.}~\bibnamefont {Xu}}, \bibinfo {author} {\bibfnamefont {J.}~\bibnamefont
  {Cai}}, \bibinfo {author} {\bibfnamefont {F.}~\bibnamefont {Hong}}, \bibinfo
  {author} {\bibfnamefont {Z.}~\bibnamefont {Fang}}, \bibinfo {author}
  {\bibfnamefont {W.}~\bibnamefont {Wang}}, \bibinfo {author} {\bibfnamefont
  {J.}~\bibnamefont {Zhu}}, \emph {et~al.},\ }\bibfield  {title} {\bibinfo
  {title} {Formation and evolution of the unexpected pbi 2 phase at the
  interface during the growth of evaporated perovskite films},\ }\href@noop {}
  {\bibfield  {journal} {\bibinfo  {journal} {Physical Chemistry Chemical
  Physics}\ }\textbf {\bibinfo {volume} {18}},\ \bibinfo {pages} {18607}
  (\bibinfo {year} {2016})}\BibitemShut {NoStop}%
\bibitem [{\citenamefont {Xia}\ \emph {et~al.}(2025)\citenamefont {Xia},
  \citenamefont {Han}, \citenamefont {Watanabe}, \citenamefont {Taniguchi},
  \citenamefont {Shan},\ and\ \citenamefont {Mak}}]{xia2025superconductivity}%
  \BibitemOpen
  \bibfield  {author} {\bibinfo {author} {\bibfnamefont {Y.}~\bibnamefont
  {Xia}}, \bibinfo {author} {\bibfnamefont {Z.}~\bibnamefont {Han}}, \bibinfo
  {author} {\bibfnamefont {K.}~\bibnamefont {Watanabe}}, \bibinfo {author}
  {\bibfnamefont {T.}~\bibnamefont {Taniguchi}}, \bibinfo {author}
  {\bibfnamefont {J.}~\bibnamefont {Shan}},\ and\ \bibinfo {author}
  {\bibfnamefont {K.~F.}\ \bibnamefont {Mak}},\ }\bibfield  {title} {\bibinfo
  {title} {Superconductivity in twisted bilayer wse2},\ }\href@noop {}
  {\bibfield  {journal} {\bibinfo  {journal} {Nature}\ }\textbf {\bibinfo
  {volume} {637}},\ \bibinfo {pages} {833} (\bibinfo {year}
  {2025})}\BibitemShut {NoStop}%
\bibitem [{\citenamefont {Guo}\ \emph {et~al.}(2025)\citenamefont {Guo},
  \citenamefont {Pack}, \citenamefont {Swann}, \citenamefont {Holtzman},
  \citenamefont {Cothrine}, \citenamefont {Watanabe}, \citenamefont
  {Taniguchi}, \citenamefont {Mandrus}, \citenamefont {Barmak}, \citenamefont
  {Hone} \emph {et~al.}}]{guo2025superconductivity}%
  \BibitemOpen
  \bibfield  {author} {\bibinfo {author} {\bibfnamefont {Y.}~\bibnamefont
  {Guo}}, \bibinfo {author} {\bibfnamefont {J.}~\bibnamefont {Pack}}, \bibinfo
  {author} {\bibfnamefont {J.}~\bibnamefont {Swann}}, \bibinfo {author}
  {\bibfnamefont {L.}~\bibnamefont {Holtzman}}, \bibinfo {author}
  {\bibfnamefont {M.}~\bibnamefont {Cothrine}}, \bibinfo {author}
  {\bibfnamefont {K.}~\bibnamefont {Watanabe}}, \bibinfo {author}
  {\bibfnamefont {T.}~\bibnamefont {Taniguchi}}, \bibinfo {author}
  {\bibfnamefont {D.~G.}\ \bibnamefont {Mandrus}}, \bibinfo {author}
  {\bibfnamefont {K.}~\bibnamefont {Barmak}}, \bibinfo {author} {\bibfnamefont
  {J.}~\bibnamefont {Hone}}, \emph {et~al.},\ }\bibfield  {title} {\bibinfo
  {title} {Superconductivity in 5.0° twisted bilayer wse2},\ }\href@noop {}
  {\bibfield  {journal} {\bibinfo  {journal} {Nature}\ }\textbf {\bibinfo
  {volume} {637}},\ \bibinfo {pages} {839} (\bibinfo {year}
  {2025})}\BibitemShut {NoStop}%
\bibitem [{\citenamefont {Posey}\ \emph {et~al.}(2024)\citenamefont {Posey},
  \citenamefont {Turkel}, \citenamefont {Rezaee}, \citenamefont {Devarakonda},
  \citenamefont {Kundu}, \citenamefont {Ong}, \citenamefont {Thinel},
  \citenamefont {Chica}, \citenamefont {Vitalone}, \citenamefont {Jing} \emph
  {et~al.}}]{posey2024two}%
  \BibitemOpen
  \bibfield  {author} {\bibinfo {author} {\bibfnamefont {V.~A.}\ \bibnamefont
  {Posey}}, \bibinfo {author} {\bibfnamefont {S.}~\bibnamefont {Turkel}},
  \bibinfo {author} {\bibfnamefont {M.}~\bibnamefont {Rezaee}}, \bibinfo
  {author} {\bibfnamefont {A.}~\bibnamefont {Devarakonda}}, \bibinfo {author}
  {\bibfnamefont {A.~K.}\ \bibnamefont {Kundu}}, \bibinfo {author}
  {\bibfnamefont {C.~S.}\ \bibnamefont {Ong}}, \bibinfo {author} {\bibfnamefont
  {M.}~\bibnamefont {Thinel}}, \bibinfo {author} {\bibfnamefont {D.~G.}\
  \bibnamefont {Chica}}, \bibinfo {author} {\bibfnamefont {R.~A.}\ \bibnamefont
  {Vitalone}}, \bibinfo {author} {\bibfnamefont {R.}~\bibnamefont {Jing}},
  \emph {et~al.},\ }\bibfield  {title} {\bibinfo {title} {Two-dimensional heavy
  fermions in the van der waals metal cesii},\ }\href@noop {} {\bibfield
  {journal} {\bibinfo  {journal} {Nature}\ }\textbf {\bibinfo {volume} {625}},\
  \bibinfo {pages} {483} (\bibinfo {year} {2024})}\BibitemShut {NoStop}%
\bibitem [{\citenamefont {Kresse}\ and\ \citenamefont
  {Furthm\"uller}(1996)}]{Kresse1996VASP}%
  \BibitemOpen
  \bibfield  {author} {\bibinfo {author} {\bibfnamefont {G.}~\bibnamefont
  {Kresse}}\ and\ \bibinfo {author} {\bibfnamefont {J.}~\bibnamefont
  {Furthm\"uller}},\ }\bibfield  {title} {\bibinfo {title} {Efficient iterative
  schemes for ab initio total-energy calculations using a plane-wave basis
  set},\ }\href {https://doi.org/10.1103/PhysRevB.54.11169} {\bibfield
  {journal} {\bibinfo  {journal} {Phys. Rev. B}\ }\textbf {\bibinfo {volume}
  {54}},\ \bibinfo {pages} {11169} (\bibinfo {year} {1996})}\BibitemShut
  {NoStop}%
\bibitem [{\citenamefont {Kresse}\ and\ \citenamefont
  {Joubert}(1999)}]{kresse1999VASP}%
  \BibitemOpen
  \bibfield  {author} {\bibinfo {author} {\bibfnamefont {G.}~\bibnamefont
  {Kresse}}\ and\ \bibinfo {author} {\bibfnamefont {D.}~\bibnamefont
  {Joubert}},\ }\bibfield  {title} {\bibinfo {title} {From ultrasoft
  pseudopotentials to the projector augmented-wave method},\ }\href
  {https://doi.org/10.1103/PhysRevB.59.1758} {\bibfield  {journal} {\bibinfo
  {journal} {Phys. Rev. B}\ }\textbf {\bibinfo {volume} {59}},\ \bibinfo
  {pages} {1758} (\bibinfo {year} {1999})}\BibitemShut {NoStop}%
\bibitem [{\citenamefont {Perdew}\ \emph {et~al.}(1996)\citenamefont {Perdew},
  \citenamefont {Burke},\ and\ \citenamefont {Ernzerhof}}]{Perdew1996}%
  \BibitemOpen
  \bibfield  {author} {\bibinfo {author} {\bibfnamefont {J.~P.}\ \bibnamefont
  {Perdew}}, \bibinfo {author} {\bibfnamefont {K.}~\bibnamefont {Burke}},\ and\
  \bibinfo {author} {\bibfnamefont {M.}~\bibnamefont {Ernzerhof}},\ }\bibfield
  {title} {\bibinfo {title} {Generalized gradient approximation made simple},\
  }\href {https://doi.org/10.1103/PhysRevLett.77.3865} {\bibfield  {journal}
  {\bibinfo  {journal} {Phys. Rev. Lett.}\ }\textbf {\bibinfo {volume} {77}},\
  \bibinfo {pages} {3865} (\bibinfo {year} {1996})}\BibitemShut {NoStop}%
\bibitem [{\citenamefont {Klime{\v{s}}}\ \emph {et~al.}(2009)\citenamefont
  {Klime{\v{s}}}, \citenamefont {Bowler},\ and\ \citenamefont
  {Michaelides}}]{klimevs2009chemical}%
  \BibitemOpen
  \bibfield  {author} {\bibinfo {author} {\bibfnamefont {J.}~\bibnamefont
  {Klime{\v{s}}}}, \bibinfo {author} {\bibfnamefont {D.~R.}\ \bibnamefont
  {Bowler}},\ and\ \bibinfo {author} {\bibfnamefont {A.}~\bibnamefont
  {Michaelides}},\ }\bibfield  {title} {\bibinfo {title} {Chemical accuracy for
  the van der waals density functional},\ }\href@noop {} {\bibfield  {journal}
  {\bibinfo  {journal} {Journal of Physics: Condensed Matter}\ }\textbf
  {\bibinfo {volume} {22}},\ \bibinfo {pages} {022201} (\bibinfo {year}
  {2009})}\BibitemShut {NoStop}%
\bibitem [{\citenamefont {Klime\ifmmode~\check{s}\else \v{s}\fi{}}\ \emph
  {et~al.}(2011)\citenamefont {Klime\ifmmode~\check{s}\else \v{s}\fi{}},
  \citenamefont {Bowler},\ and\ \citenamefont {Michaelides}}]{klimves2011}%
  \BibitemOpen
  \bibfield  {author} {\bibinfo {author} {\bibfnamefont {J.~c.~v.}\
  \bibnamefont {Klime\ifmmode~\check{s}\else \v{s}\fi{}}}, \bibinfo {author}
  {\bibfnamefont {D.~R.}\ \bibnamefont {Bowler}},\ and\ \bibinfo {author}
  {\bibfnamefont {A.}~\bibnamefont {Michaelides}},\ }\bibfield  {title}
  {\bibinfo {title} {Van der waals density functionals applied to solids},\
  }\href {https://doi.org/10.1103/PhysRevB.83.195131} {\bibfield  {journal}
  {\bibinfo  {journal} {Phys. Rev. B}\ }\textbf {\bibinfo {volume} {83}},\
  \bibinfo {pages} {195131} (\bibinfo {year} {2011})}\BibitemShut {NoStop}%
\bibitem [{\citenamefont {Dion}\ \emph {et~al.}(2004)\citenamefont {Dion},
  \citenamefont {Rydberg}, \citenamefont {Schr\"oder}, \citenamefont
  {Langreth},\ and\ \citenamefont {Lundqvist}}]{Dion2004}%
  \BibitemOpen
  \bibfield  {author} {\bibinfo {author} {\bibfnamefont {M.}~\bibnamefont
  {Dion}}, \bibinfo {author} {\bibfnamefont {H.}~\bibnamefont {Rydberg}},
  \bibinfo {author} {\bibfnamefont {E.}~\bibnamefont {Schr\"oder}}, \bibinfo
  {author} {\bibfnamefont {D.~C.}\ \bibnamefont {Langreth}},\ and\ \bibinfo
  {author} {\bibfnamefont {B.~I.}\ \bibnamefont {Lundqvist}},\ }\bibfield
  {title} {\bibinfo {title} {Van der waals density functional for general
  geometries},\ }\href {https://doi.org/10.1103/PhysRevLett.92.246401}
  {\bibfield  {journal} {\bibinfo  {journal} {Phys. Rev. Lett.}\ }\textbf
  {\bibinfo {volume} {92}},\ \bibinfo {pages} {246401} (\bibinfo {year}
  {2004})}\BibitemShut {NoStop}%
\bibitem [{\citenamefont {Bl\"ochl}(1994)}]{Blochl94}%
  \BibitemOpen
  \bibfield  {author} {\bibinfo {author} {\bibfnamefont {P.~E.}\ \bibnamefont
  {Bl\"ochl}},\ }\bibfield  {title} {\bibinfo {title} {Projector augmented-wave
  method},\ }\href {https://doi.org/10.1103/PhysRevB.50.17953} {\bibfield
  {journal} {\bibinfo  {journal} {Phys. Rev. B}\ }\textbf {\bibinfo {volume}
  {50}},\ \bibinfo {pages} {17953} (\bibinfo {year} {1994})}\BibitemShut
  {NoStop}%
\end{thebibliography}%
\section{Methods}
\subsection{First-principles computations}
All first-principles calculations based on the density-functional theory are performed using the Vienna \textit{ab initio} simulation package~\cite{Kresse1996VASP,kresse1999VASP}, and the exchange-correlation potentials use the Perdew-Burke-Ernzerhof (PBE) parametrization of the generalized gradient approximation~\cite{Perdew1996}.
Automatic grid generation was employed to properly sample the Brillouin zone, such as the spacing (in-plane) between $k$-points was $0.15 \AA^{-1}$. A plane-wave cutoff of $E=450\textrm{eV}$ was applied in all calculations. Structural relaxation for $AA$ stacked and shifted configuration was carried out without the effects of SOC. Relaxation was carried out, in the AA case, until the forces on individual atoms $|\mathbf{f}_i| < 5 \rm{meV}/\AA$. When creating the shifted configurations, the in-plane motion of the atoms was clamped, allowing relaxation only in the out-of-plane, $z$ direction, with the same convergence criterion. Two-body vdW dispersive corrections were considered for relaxation using the method of Klimes \cite{klimevs2009chemical,klimves2011} and Dion \textit{et al.} \cite{Dion2004}.
In the generation of Wannier Hamiltonians, non-collinear SOC was included and the system was initialized in a non-magnetic state. 

In order to automate the process, we Wannierize the model according to the valence shell defined in the PAW pseudopotentials \cite{Blochl94}. 

\subsection{Twisted bilayer Hamiltonian construction}
As stated in the main body, the Hamiltonian for the twisted bilayer is constructed via the PyBinding software package\cite{dean_moldovan_2020_4010216}. We begin by importing the WTB model for the untwisted bilayer and removing all interlayer couplings. This creates a block diagonal Hamiltonian where the two-blocks are given by the WTB for each layer, with no restriction on hopping distance or magnitude. Next, using Pybinding's finite size module, a supercell of the bilayer is created from which a circle of radius $R=50\AA$ is cut as shown in Fig.~\eqref{fig:Fig1}. A function is then introduced allowing the bottom layer to be twisted to an arbitrary angle, $\theta$, relative to the top layer. After imposing the desired twist, we create a register of all atomic pairs in distinct layers within a distance set by $2\times d_{inter}$, where $d_{inter}$ is the average interlayer spacing. For each pair in this registry, we loop through all possible interlayer hopping matrix elements, corresponding to orbital/spin pairs. If a random forest model has been trained for a given pair, indicating a finite interlayer hopping strength was found between the pair during the ``local-configuration'' computations, a value of the interlayer hopping is generated by supplying the distance and angle between the pair to the random forest model and incorporated in the Hamiltonian. As the bilayers studied in this work are selected from a database of easily exfoliable two-dimensional materials~\cite{mounet2018two,campi2023expansion}, they are primarily Van der Waals systems. As a result, we choose to leave the intralayer Hamiltonian fixed for all twist angles, adjusting only the interlayer terms.

\subsection{Estimating the Density of states}
Due to having open boundary conditions we estimate the bulk DOS through an average over sites in the middle of the sample.
The local density of states is defined as
$\rho_{\bm{r},\tau,\sigma}(E)=\sum_{n}|\langle n|\bm r, \sigma, \tau\rangle|^2\delta(E-E_{n})$
where $i,\tau,\sigma$ correspond to site, orbital and spin degrees of freedom, restricting the sum over $i$ to all sites within a $10 \AA$ radius of the sample center.

\subsection{Performance of machine learning model} 
As stated in Sec.~\eqref{sec:2}, interlayer hopping matrix elements for the twisted bilayer Hamiltonians constructed in this work are produced using an ensemble of random forest machine learning (ML) models. We require an ensemble of models because a new model is trained for each atom,orbital, spin pair located in distinct layers. To limit the computational expense, we only train a ML model of a given pair supports at least one hopping element greater than $0.01eV$ among all generated tight-binding models produced in the local-configuration computations using DFT~\cite{kresse1999VASP,Kresse1996VASP} and Wannier90~\cite{Pizzi2020}.
\allowdisplaybreaks
\begin{figure}
    \centering
    \includegraphics[width=8cm]{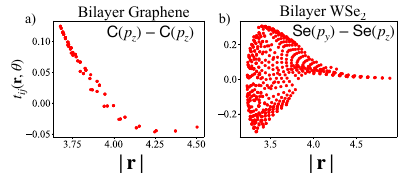}
    \caption{\textbf{Fitting interlayer hoppings :} Interlayer hopping matrix elements extracted from local stacking configuration computations as a function of distance, $|\mathbf{r}|$ between (a) $p_{z}$ orbitals in distinct layers of bilayer graphene and (b) $p_{z}$ and $p_{y}$ orbitals of Se in distinct layers of bilayer WSe$_{2}$. The spread in data points indicates the $\theta$ dependence of eq.~\eqref{eq:fithop}. Data illustrates increased complexity of spin-orbit coupled compounds such as WSe$_{2}$. Imaginary contribution to hoppings is neglected for WSe$_{2}$ as it falls below the threshold of $0.01eV$.}
    \label{fig:InterlayerHoppings}
\end{figure}

Here we make the role of the ML networks concrete. In Ref.~\cite{Fang2016} it is given that for Wannier centers in distinct monolayers, monolayer 1 and 2, brought into contact, if a given monolayer and Wannier center admit an $N$-fold rotational symmetry, the interlayer hoppings which compose, $\mathcal{T}(\theta)$ in eq.~\eqref{eq:BilayerHam}, can be fit as a function of the relative vector between the centers, $\mathbf{r}$, and the angles $\theta_{i=1,2}$, where $\theta_{i}$ is the angle relative to $\mathbf{r}$ necessary to determine the orientation of monolayer $i$. Namely, the interlayer hopping takes the form, 

\begin{equation}\label{eq:fithop}
    t_{ij}(\mathbf{r},\theta_{1},\theta_{2})=\sum_{m_{1},m_{2}=-\infty}^{+\infty}f_{m_{1},m_{2} }(\mathbf{r})e^{im_{1}\theta_{1}/N_{1}+im_{2}\theta_{2}/N_{2}}.
\end{equation}
As we focus only on homo-bilayers and always leave the bottom layer fixed, twisting the top layer, this function can be reduced, fixing $m_{2}=0$, to a function of the relative vector $r$ and a single angle $\theta=\theta_{1}$, which provides the relative twist of the top layer. 
\par 
We now examine the behavior of this function in bilayer graphene and bilayer WSe$_{2}$ by plotting the interlayer hoppings as a function of their distance, $|\mathbf{r}|$, extracted from the WTBs constructed in the local stacking configuration protocol. This data is displayed in Fig.~\eqref{fig:InterlayerHoppings}(a)  for a pair of $p_{z}$ orbitals in distinct layers of graphene. In Fig.~\eqref{fig:InterlayerHoppings}(b) the hopping between an Se $p_{y}$ and an Se $p_{z}$ orbital in distinct layers of WSe$_{2}$ is shown. The distribution for bilayer graphene has been studied in Ref. \cite{Fang2016}, where it is noted that the spread in the data as a function of $|\mathbf{r}|$ is due to $\theta$ dependence. While the $\theta$ dependence is minimal for graphene, it is important to emphasize the increased complexity of this dependence in a TMD such as WSe$_{2}$ where distinct orbitals couple. The ML models trained in this work are designed to automate fitting of this complex function in an efficient and accurate manner, amenable to high-throughput screenings.

\begin{figure}
    \centering
    \includegraphics[width=8cm]{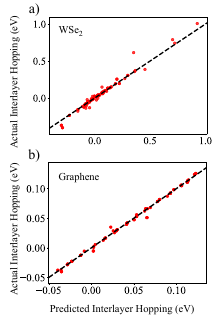}
    \caption{\textbf{Machine learning model performance: }Performance of random forest machine learning models on test set of interlayer couplings for local stacking configurations of (a) WSe$_{2}$ and (b) graphene. The results are aggregated for 18 and 4 trained random forest models for WSe$_{2}$ and graphene, respectively.}
    \label{fig:ModelPerformance}
\end{figure}
\par 
These regression models take as input the relative vector between the sites $\mathbf{r}=\{r_{x},r_{y},r_{z}\}$, broken down into two components, magnitude, $|\mathbf{r}|$, and in-plane angle, $\phi=\tan^{-1}(r_{y}/r_{x})$, shown schematically in Fig.~\eqref{fig:input_data}. As stated in the main body, we consider only a single orbital/spin pair at a time. We can thus invoke the periodicity of $\phi$ to fix $\phi=0$ for the untwisted unit cell. It is then possible to define $\phi =\theta_{1}-\theta_{2}=\theta$. The model output is then the hopping matrix element. To demonstrate the performance of this approach, we display in Fig.~\eqref{fig:ModelPerformance}, the accuracy of the models for two materials of great interest to the community, bilayer graphene and bilayer WSe$_{2}$. The random forest models are trained using the mean absolute error (MAE) as the loss function. The results on the test set for WSe$_{2}$ and graphene when averaging over all interlayer pairs and considering a $90\%/10\%$ train/test split of the data is shown in Fig.~\eqref{fig:ModelPerformance}(A) and (B) respectively. The MAE is $0.021 \pm 0.002 eV$ for WSe$_{2}$ and $0.0017 \pm 0.0003eV$ for graphene. This is impressive accuracy given the limited dataset and the fact that training of the models, 18 in the case of WSe$_{2}$ and 4 in the case of graphene, took an average of $\sim 0.5s$ of training time per model on a single laptop CPU. This fact underscores the power of the model to deliver state of the art accuracy without the need for computationally expensive ab initio computations of the full twisted unit cell, or complex graph neural networks requiring high-performance GPUs. This is a vital framework for extending high-throughput workflows to Moire systems.

\end{document}